\documentclass[twocolumn,showpacs,preprintnumbers,amsmath,amssymb]{revtex4}

\usepackage{graphicx}
\usepackage{dcolumn}
\usepackage{bm}

\begin{document}
\begin{widetext}
\thispagestyle{empty}
\begin{Large}
\textbf{DEUTSCHES ELEKTRONEN-SYNCHROTRON}

\textbf{\large{in der HELMHOLTZ-GEMEINSCHAFT}\\}
\end{Large}

DESY 04-112

July 2004

\begin{eqnarray}
\nonumber &&\cr \nonumber && \cr \nonumber &&\cr
\end{eqnarray}
\begin{eqnarray}
\nonumber
\end{eqnarray}
\begin{center}
\begin{Large}
\textbf{Theory of space-charge waves on gradient-profile
relativistic electron beam: an analysis in propagating eigenmodes}
\end{Large}
\begin{eqnarray}
\nonumber &&\cr \nonumber && \cr
\end{eqnarray}

\begin{large}
Gianluca Geloni, Evgeni Saldin, Evgeni Schneidmiller and Mikhail
Yurkov
\end{large}
\textsl{\\Deutsches Elektronen-Synchrotron DESY, Hamburg}
\begin{eqnarray}
\nonumber
\end{eqnarray}
\begin{eqnarray}
\nonumber
\end{eqnarray}
\begin{eqnarray}
\nonumber
\end{eqnarray}
\begin{eqnarray}
\nonumber
\end{eqnarray}
\begin{eqnarray}
\nonumber
\end{eqnarray}
\begin{eqnarray}
\nonumber
\end{eqnarray}
\begin{eqnarray}
\nonumber
\end{eqnarray}
\begin{eqnarray}
\nonumber
\end{eqnarray}
\begin{eqnarray}
\nonumber
\end{eqnarray}
\begin{eqnarray}
\nonumber
\end{eqnarray}
\begin{eqnarray}
\nonumber
\end{eqnarray}
ISSN 0418-9833
\begin{eqnarray}
\nonumber
\end{eqnarray}
\begin{large}
\textbf{NOTKESTRASSE 85 - 22607 HAMBURG}
\end{large}
\end{center}
\end{widetext}


\title{Theory of space-charge waves on gradient-profile relativistic electron beam:
\\ an analysis in propagating eigenmodes}

\author{Gianluca Geloni}
 \email{gianluca.aldo.geloni@desy.de}
 \altaffiliation[Also at ]{Department of Applied Physics, Technische Universiteit Eindhoven, The Netherlands}
\author{Evgeni Saldin}%
\author{Evgeni Schneidmiller}
\author{Mikhail Yurkov}
 \altaffiliation[Also at ]{Joint Institute for Nuclear Research, Dubna, Moscow Region, Russia}
\affiliation{%
Deutsches Elektronen-Synchrotron, Notkestrasse 85, 22607 Hamburg,
\\ Germany}

\date{\today}

\begin{abstract}
We developed an exact analytical treatment for space-charge waves
within a relativistic electron beam in terms of (self-reproducing)
propagating eigenmodes. This result is of obvious theoretical
relevance as it constitutes one of the few exact solution for the
evolution of charged particles under the action of
self-interactions. It also has clear numerical applications in
particle accelerator physics where it can be used as the first
self-consistent benchmark for space-charge simulation programs.
Today our work is of practical relevance in FEL technology in
relation with all those schemes where an optically modulated
electron beam is needed and with the study of longitudinal
space-charge instabilities in magnetic bunch compressors.
\end{abstract}

\maketitle

\section{\label{sec:intro} Introduction}

The evolution problem for a collection of charges under the action
of their own fields when certain initial conditions are given is,
in general, a formidable one. Numerical methods are often the only
way to obtain an approximate solution, while there are only a few
cases in which finding an exact treatment is possible.

In this paper we report a fully self-consistent solution to one of
these problems, namely the evolution of a relativistic electron
beam under the action of its own fields in the (longitudinal)
direction of motion. The problem of longitudinal space-charge
oscillations has been, so far, solved only from an
electrodynamical viewpoint \cite{ROSE}, or using limited
one-dimensional models \cite{PLA2}. On the contrary, in our
derivation the beam, which is assumed infinitely long in the
longitudinal direction, is accounted for any given radial
dependence of the particle distribution function.

An initial condition is set so that the beam is considered
initially modulated in energy and density at a given wavelength.
When the amplitude of the modulation is small enough the evolution
equation can be linearized. An exact solution can be found in
terms of an expansion in (self-reproducing) propagating
eigenmodes.

Our findings are, in first instance, of theoretical importance
since they constitute one of the few exact solutions known up to
date to the problem of particles evolving under the action of
their own fields.

Next to theoretical and academical interest of our study we want
to emphasize here its current relevance to applied physics and
technology. For example, particle accelerator physics in general
and FEL physics in particular make large use of simulation codes
(see for instance \cite{SIMS,TDRD,CDRS}) in order to obtain the
influence of space-charge fields on the beam behavior. Yet, these
codes are benchmarked against exact solutions of the
electrodynamical problem alone (i.e. solutions of Poisson
equation) and only recently \cite{PLA2} partial attempts have been
made to benchmark them against some analytical model accounting
for the system evolution. However, such attempts are based on
one-dimensional theory which can only give some incomplete result.
On the contrary, we claim that our findings can be used as a
standard benchmark for any space-charge code from now on.

To give some up-to-date example of practical applications
(besides, again, theoretical and numerical importance) we wish to
underline that our results are of relevance to an entire class of
problems arising in state-of-the-art FEL technology. In fact,
several applications rely on feeding (optically) modulated
electron beams into an FEL. For instance, optical seeding is a
common technique for harmonic generation \cite{MOD0}. Moreover
pump-probe schemes have been proposed that couple the laser pulse
out of an X-ray FEL with an optical laser. To solve the problem of
synchronization between the two lasers a method has been proposed,
which makes use of a single optically modulated electron beam in
order to generate both pulses. This proposal relies on the passage
of this electron beam through an X-ray FEL and an optically tuned
FEL \cite{MOD1}: given the parameters of the system, plasma
oscillations turn out to be a relevant effect to be accounted for.
It is also worth to mention, as another example, the relevance of
plasma oscillation theory in the understanding of practical issues
like longitudinal space-charge instabilities in high-brightness
linear accelerators. High-frequency components of the bunch
current spectrum (at wavelengths much shorter than the bunch
length) can induce, through self-interaction, energy modulation
within the electron beam which is then converted into density
modulation when the beam passes through a magnetic compression
chicane, thus leading to beam microbunching and break up. When
longitudinal space-charge is the self-interaction driving the
instability (see \cite{MICR}), one is interested to know with the
best possible accuracy how plasma oscillations modulate the bunch,
in energy and density, before the compression chicane.

Our calculations can be applied to these cases directly or in
support to macroparticle simulations thus providing outcomes of
immediate practical importance.

Throughout this paper we will make use of $cgs$ units. Our work is
organized as follows. Next to this Introduction, in Section
\ref{sec:theory}, we pose the problem in the form of an
integro-differential equation which makes use of naturally
normalized quantities. In Section \ref{sub:axis} we present our
main theoretical results. In the following Section \ref{sub:test}
we describe some applications and important exemplifications of
the obtained results, including the role of the initial condition.
Finally, in Section \ref{sec:concl} we come to conclusions.

\section{\label{sec:theory} Theory}

\subsection{\label{sub:hami} Position of the problem}

We are interested in developing a theory to describe longitudinal
plasma waves in a relativistic electron beam. In order to do so we
have to find a way to translate in mathematical terms the idea of
dealing with the longitudinal dynamics only, while keeping intact
the general three-dimensional description of the system
(electromagnetic fields and particle distribution).

Immediately related with the development of a theory, is its range
of applicability. The idea of selecting longitudinal dynamics
translates from a physical viewpoint in the assumption that the
longitudinal space-charge fields, alone, describe the system
evolution. This corresponds to a situation with physical
parameters tuned in such a way that, on the time-scale of a
longitudinal plasma oscillation, transverse dynamics does not play
a role.

One may, of course, devise different methods to select the
longitudinal dynamics alone: actually he will come up with some
ideal approximation of what real focusing systems in modern linear
accelerators do, provided that the $\beta$-functions are large
enough with respect to the plasma wavelength.

Whatever the practical or ideal mean chosen to select longitudinal
dynamics such a choice translates, from a mathematical viewpoint,
in the choice of dynamical variables along the direction of motion
only, while transverse coordinates enter purely as parameters in
the description of the fields and of the particle distribution.

Our beam is initially modulated at some wavelength $\lambda_m$, in
density and energy. This is no restriction because, as said in
Section \ref{sec:intro}, the modulation amplitude is considered
small enough so that we can linearize the evolution equation.
Then, a Fourier analysis of any perturbation in energy and density
is customary. Once the modulation wavelength $\lambda_m$ has been
fixed, it is natural to define the phase $\psi = \omega_m
\left(z/v_z(\mathcal{E}_0) - t\right)$, where $v_z(\mathcal{E}_0)
\sim c$ is the longitudinal electron velocity at the nominal beam
kinetic energy $\mathcal{E}_0 = (\gamma-1)m c^2$, $\omega_m = 2
\pi v_z/\lambda_m$, $t$ is the time and $z$ the longitudinal
abscissa. Upon this, and after what has been said about the choice
of longitudinal dynamics, it is appropriate to operate in
energy-phase variables $(P,\psi)$, $P$ being the deviation from
the nominal energy.

In this spirit the total derivative of the phase $\psi$ is given
by:

\begin{equation}
{d\psi \over{dz}} = {\partial \psi\over{\partial z}}+ {\partial
\psi\over{\partial t}} {d t\over{d z}}=
{\omega_m\over{v_z(\mathcal{E}_0)}}-
{\omega_m\over{v_z(\mathcal{E})}}\label{psieq}~.
\end{equation}
Now, if we assume that the particle energy is not significantly
different from the nominal energy  we can expand $v(\mathcal{E})$
in $\mathcal{E}$ around $\mathcal{E}_0$. Keeping up to second
order terms in $\mathcal{E}$ and using the definition of $P$ we
find

\begin{equation}
{d\psi \over{dz}} = {\omega_m
P\over{c\gamma_z^2\mathcal{E}_0}}\label{psieq2}~,
\end{equation}
where we took advantage of the fact that
$(dv_z/d\mathcal{E})\mid_{\mathcal{E}=\mathcal{E}_0} \simeq
c/(\gamma_z^2\mathcal{E}_0$), where $\gamma_z =
(1-v_z(\mathcal{E}_0)^2/c^2)^{-1/2}$. Note that, here, we
distinguish from the very beginning between $\gamma$ and
$\gamma_z$ (or $v$ and $v_z$). In fact our theory can be applied
to the case of an electron beam in vacuum as well as to the case
of a beam under the action of external electromagnetic fields, for
example in an undulator: in the first situation $\gamma=\gamma_z$,
strictly, while in the latter they obviously have different
values.

The full derivative of $P$ is simply given by
\begin{equation}
{d P \over{dz}} = -e E_z\label{Peq}~,
\end{equation}
where $E_z(z,\psi)$ is the space-charge field in the $z$
direction. Eq. (\ref{psieq2}) and Eq. (\ref{Peq}) are the equation
of motion for our system and they can be interpreted as Hamilton
canonical equations corresponding to the Hamiltonian
$H(\psi,P,z)$:

\begin{equation}
H = e \int{d\psi E_z} + {\omega_m\over{2 c \gamma_z^2}}
{P^2\over{\mathcal{E}_0}}~. \label{hfin}
\end{equation}
In this sense, Eq. (\ref{hfin}), alone, defines our theory. The
bunch density distribution will be then represented by the density
$f = f(\psi,P,z;\bm{r_\bot})$, where the semicolon separates
dynamical variables from parameters and it will be subjected to
the (Vlasov) evolution equation

\begin{equation}
{\partial f\over{\partial z}} + {\partial H\over{\partial P}}
{\partial f\over{\partial \psi}} - {\partial H\over{\partial
\psi}}
 {\partial f\over{\partial P}} = 0~
\label{vla}
\end{equation}
with appropriate initial condition at $z=0$. Here we are
interested in a beam initially modulated in energy and density;
moreover, as said in Section \ref{sec:intro}, the modulation must
be small enough to ensure that linearization of Eq. (\ref{vla}) is
possible. Thus we will take $f(\psi,P,z;\bm{r_\bot})_{\mid_{z=0}}
= f_0(P;\bm{r_\bot}) + f_1(\psi,P,z;\bm{r_\bot})_{\mid_{z=0}}$.
$f_0$ is a so called unperturbed solution of the evolution
equation Eq. (\ref{vla}), and it does not depend on $z$, while
$f_1$ is known as the perturbation; it is understood that, in
order to be in the linear regime, $f_1 \ll f_0$  for any value of
dynamical variables or parameters. In the following we assume that
the dynamical variable $P$ and the parameter $\bm{r_\bot}$ are
separable in $f_0$ so that we may write $f_0(P;\bm{r_\bot}) =
n_0(\bm{r_\bot}) F(P)$, where the local energy spread function
$F(P)$ is considered normalized to unity. The initial modulation
can be written as a sum of density and energy modulation terms:
$f_1(\psi,P,z;\bm{r_\bot})_\mid{_{z=0}} =
f_{1d}(\psi,P;\bm{r_\bot})+f_{1e}(\psi,P;\bm{r_\bot})$.

On the one hand $f_{1d}$ is responsible for a pure density
modulation and can be written as

\begin{equation}
f_{1d}(\psi,P,z;\bm{r_\bot}) = a_{1d}(\bm{r_\bot}) F(P)
\cos(\psi)~, \label{densd}
\end{equation}
where we set to zero an unessential, initial modulation phase.

On the other $f_{1e}$ is responsible for a pure energy modulation
and can be assumed to be

\begin{equation}
f_{1e}(\psi,P,z;\bm{r_\bot}) = a_{1e}(\bm{r_\bot}) {dF\over{dP}}
\cos(\psi+\psi_0)~, \label{dense}
\end{equation}
where $\psi_0$ is an initial (relative) phase between density and
energy modulation. Finally it is convenient to define complex
quantities $\tilde{f}_{1d}=a_{1d} F$, and $\tilde{f}_{1e}=a_{1e}
({dF}/{dP}) e^{i\psi_0}$ so that ${f}_{1_\mid{_{z=0}}} =
(\tilde{f}_{1d}+\tilde{f}_{1e})e^{i\psi} + CC$. In the linear
regime, then, one can write
$f_{1}(\psi,P,z;\bm{r_\bot})=\tilde{f}_1(P,z;\bm{r_\bot})e^{i\psi}+
CC$.  Further definition of $\tilde{E}_z =\tilde{E}
(z;\bm{r_\bot})$ in such a way that $E_z = \tilde{E}_z e^{i \psi}
+ \tilde{E}_z^* e^{-i \psi}$ allows one to write down the Vlasov
equation, Eq. (\ref{vla}), linearized in $\tilde{f}_1$:

\begin{equation}
{\partial \tilde{f}_1\over{\partial z}} + i  {\omega_m P\over{c
\gamma_z^2 \mathcal{E}_0}} \tilde{f}_1 - e \tilde{E}_z {\partial
f_0\over{\partial P}} = 0~. \label{vlalin}
\end{equation}
Eq. (\ref{vlalin}) is far from being the final form of the
evolution equation, since we still have to couple it with Maxwell
equations, which constitute the electrodynamical part of the
problem. However an integral representation of $\tilde{f}_1$ can
be given at this stage:

\begin{equation}
\tilde{f}_1 = \tilde{f}_{1_{\mid_{z=0}}} e^{-{i \omega_m P z\over
{c \gamma_z^2 \mathcal{E}_0}}} + e n_0 {d F\over{d P}} \int_0^z{
dz' \tilde{E}_z e^{i {\omega_m P\over{c \gamma_z^2 \mathcal{E}_0}}
(z'-z) }}~. \label{integrapr}
\end{equation}
Let us now introduce the longitudinal current density
$j_z(z;\bm{r_\bot}) =  -j_0(\bm{r_\bot}) + \tilde{j}_1 e^{i \psi}
+ \tilde{j}_1^* e^{-i \psi}$, where $j_0(\bm{r_\bot}) \simeq  e c
n_0(\bm{r_\bot})$ and $\tilde{j}_1 \simeq -e c
\int_{-\infty}^{\infty} dP \tilde{f}_1$. Eq. (\ref{integrapr}) can
be integrated in $P$ thus giving

\begin{eqnarray}
\tilde{j}_1 = -e c\int_{-\infty}^{\infty} dP \left(a_{1d}F +
a_{1e}{dF\over{dP}}e^{i\psi_0}\right) e^{-i {\omega_m P z\over{c
\gamma_z^2 \mathcal{E}_0}}} && \cr - e j_0 \int_0^z dz'
\left[\tilde{E}_z \int_{-\infty}^{\infty} dP {d F\over{d P}} e^{i
{\omega_m P\over{c \gamma_z^2 \mathcal{E}_0}} (z'-z)} \right]~.
\label{integrapr2}
\end{eqnarray}
The next step is to present the equation for the electric field
$\tilde{E}_z$ which, coupled with Eq. (\ref{integrapr2}), will
describe the system evolution in a self-consistent way.

We start with the inhomogeneous Maxwell equation for the
z-component of the electric field

\begin{equation}
\nabla^2 E_z - {1\over{c^2}} {\partial^2 E_z\over{\partial t^2}} =
4 \pi {\partial \rho_e\over{\partial z}} +
{4\pi\over{c^2}}{\partial j_z\over{\partial t}}~,\label{max1}
\end{equation}
where $\rho_e$ is the electron charge density. Remembering the
definition of complex quantities $\tilde{E}_z$ and $\tilde{j}_1$
and accounting for the fact that $\rho_e \simeq j_z/v_z$ one can
rewrite Eq. (\ref{max1}) as

\begin{eqnarray}
\nabla_\bot^2 \left(\tilde{E}_z e^{i\psi}\right)+{\partial^2
\tilde{E}_z e^{i\psi}\over{\partial z^2}}- {1\over{c^2}}
{\partial^2 \tilde{E}_z e^{i\psi}\over{\partial t^2}} &&\cr= {4
\pi\over{v_z}} {\partial \tilde{j}_1 e^{i\psi}\over{\partial z}} +
{4\pi\over{c^2}}{\partial \tilde{j}_1 e^{i\psi}\over{\partial
t}}~,\label{max2}
\end{eqnarray}
where $\nabla_\bot^2$ is the Laplacian operator over transverse
coordinates. Explicit calculations of partial derivatives with
respect to $t$ and $z$ give

\begin{equation}
{\partial^2 \tilde{E}_z e^{i\psi}\over{\partial z^2}} = \left(
{\partial^2 \tilde{E}_z \over{\partial z^2}}  + 2 i
{\omega_m\over{v_z}} {\partial \tilde{E}_z \over{\partial z}}
-{\omega_m^2\over{v_z^2}} \tilde{E}_z \right)e^{i\psi}~,
\label{der1}
\end{equation}
\begin{equation}
{\partial^2 \tilde{E}_z e^{i\psi}\over{\partial t^2}} =
-\omega_m^2 \tilde{E}_z e^{i\psi}~,\label{der2}
\end{equation}
\begin{equation}
{\partial \tilde{j}_1 e^{i\psi}\over{\partial z}} = {\partial
\tilde{j}_1 \over{\partial
z}}e^{i\psi}+i{\omega_m\over{v_z}}\tilde{j}_1
e^{i\psi}~,\label{der3}
\end{equation}
\begin{equation}
{\partial \tilde{j}_1 e^{i\psi}\over{\partial t}} = -i \omega_m
\tilde{j}_1 e^{i\psi}~.\label{der4}
\end{equation}
Substitution back into Eq. (\ref{max2}) yields ($v_z\simeq c$):

\begin{eqnarray}
\nabla_\bot^2 \tilde{E}_z + {\partial^2 \tilde{E}_z \over{\partial
z^2}}  + 2 i {\omega_m\over{v_z}} {\partial \tilde{E}_z
\over{\partial z}}  - {\omega_m^2 \tilde{E}_z\over{\gamma_z^2
c^2}} &&\cr= {4 \pi\over{v_z}} {\partial \tilde{j}_1
\over{\partial z}}+{4 \pi i \omega_m \over{\gamma_z^2 c^2
}}\tilde{j}_1~.\label{max3a}
\end{eqnarray}
It is reasonable to assume that the envelope of fields and
currents vary slowly enough over the $z$ coordinate, in order to
neglect first and second derivatives with respect to $z$ in Eq.
(\ref{max3a}). Mathematically, this corresponds to the
requirements

\begin{equation}
{\left|\partial \tilde{E}_z \over{\partial z}\right|\ll{k_m
\over{2 \gamma_z^2}} {\left|\tilde{E}_z\right|}}~,
\label{assumpsempl}
\end{equation}
\begin{equation}
{\left|\partial^2 \tilde{E}_z \over{\partial z^2}\right| \ll{k^2_m
\over{\gamma_z^2}}{\left|\tilde{E}_z\right|}}~
\label{assumpsempl2}
\end{equation}
and
\begin{equation}
{\left|\partial \tilde{j}_1 \over{\partial z}\right|\ll{k_m
\over{\gamma_z^2}}{\left|\tilde{j}_1\right|}}~,
\label{assumpsempl3}
\end{equation}
where we introduced the wave number $k_m = 2\pi/\lambda_m$.  We
note that $2\gamma^2_z\lambda_m$ is, roughly, the field formation
length: by imposing conditions (\ref{assumpsempl}),
(\ref{assumpsempl2}) and (\ref{assumpsempl3}) we are requiring
that the characteristic lengths of variation for current, field
and its derivative are much longer than the field formation length
(actually condition (\ref{assumpsempl2}) over the variation of the
field derivative already included in condition
(\ref{assumpsempl}), since $\gamma_z^2 \gg 1$): this simply means
that we can neglect retardation effects or, in other words, that
the fields are known at a certain time, when the charge
distribution are known at the same time. This assumption is not a
restriction and it is verified in all cases of practical interest.
Then Eq. (\ref{max3a}) is simplified to

\begin{equation}
\nabla_\bot^2 \tilde{E}_z  - {\omega_m^2
\tilde{E}_z\over{\gamma_z^2 c^2}} = {4 \pi i
\omega_m\over{\gamma_z^2 c^2}} \tilde{j}_1 ~,\label{max3}
\end{equation}
which forms, together with Eq. (\ref{integrapr2}), a
self-consistent description for our system.

Similarity techniques can be now used in order to obtain a
dimensionless version of Eq. (\ref{integrapr2}) and Eq.
(\ref{max3}). First note that the dependence of $j_0$ on the
transverse coordinates can be expressed, in all generality, as

\begin{equation}
j_0 = I_0 S_0(\bm{r_\bot}/r_0) \left[\int S_0(\bm{r_\bot}/r_0)
d\bm{r_\bot}\right]^{-1}~, \label{j0radial}
\end{equation}
where $I_0$ is the beam current, $r_0$ the transverse profile
parameter (i.e. the typical transverse size of the beam), $S_0$
the transverse profile function of the beam and the integral in
$d\bm{r_\bot}$ is calculated over all the transverse plane.
Furthermore it is understood that the normalization of $S_0$ is
chosen is such a way that $S_0(\bm{0})=1$. It is then customary to
introduce the current density parameter $J_0 = I_0 \left[\int
S(\bm{r_\bot}/r_0) d\bm{r_\bot}\right]^{-1}$ so that we can define
quite naturally the dimensionless current densities $\hat{j}_0 =
j_0/J_0 \equiv S_0(\bm{r_\bot}/r_0)$ and $\hat{j}_1 =
\tilde{j}_1/J_0$. It follows from Eq. (\ref{max3}) that the
electric field should be normalized to $E_0 = 4 \pi J_0/\omega_m$,
which suggests the definition $\hat{E}_z=\tilde{E}_z/E_0$. For the
normalization of the transverse coordinates we use ${\bf{\hat{r}}}
= \bm{r_\bot}/r_0$ so that  one is naturally guided by Eq.
(\ref{max3}) to introduce the transverse size parameter

\begin{equation}
q = k_m r_0/\gamma_z ~.\label{qpar}
\end{equation}
Eq. (\ref{max3}) can now be written in its final dimensionless
form:

\begin{equation}
\hat{\nabla}_\bot^2 \hat{E}_z  - q^2 \hat{E}_z = i q^2 \hat{j}_1
~,\label{max4}
\end{equation}
where $\hat{\nabla}_\bot^2$ is the Laplacian operator with respect
to normalized transverse coordinates.

By analyzing Eq. (\ref{integrapr2}) and using the normalized
quantities defined above we recover a dimensionless variable $\hat
z = \Lambda_P z$, where $\Lambda_P$ is given by

\begin{equation}
\Lambda_P = \left[4 I/(I_A r_0^2 \gamma \gamma_z^2) \right]^{1/2}
\label{lmbdapar} ~,
\end{equation}
$I_A=m c^3/e$ being the Alfven current.

Using Eq. (\ref{lmbdapar}) and looking now at the exponential
factors in Eq. (\ref{integrapr2}) it is straightforward to
introduce the dimensionless energy deviation $\hat{P} = P/(\rho
\mathcal{E}_0)$, $\rho$ being defined by

\begin{equation}
\rho = {\Lambda_p \gamma_z^2 \over{ k_m}}~. \label{rhopar}
\end{equation}
From the definition of $\hat{P}$ it follows immediately that the
factor $\rho \mathcal{E}_0$ is a natural measure for energy
deviations. The rms energy spread $\langle (\Delta
\mathcal{E})^2\rangle$ can be measured by the dimensionless
parameter

\begin{equation}
\hat{\Lambda}_T^2 =  {\langle (\Delta
\mathcal{E})^2\rangle\over{\rho^2
\mathcal{E}_0^2}}~.\label{ensprpar}
\end{equation}

The local energy spread distribution $F(P)$ was defined as
normalized to unity, so that it is customary to introduce
$\hat{F}(\hat{P})$ as the distribution function in the reduced
momentum $\hat{P}$ also normalized to unity. For example, when the
energy spread is a gaussian we have $\hat{F}(\hat{P}) = (2\pi
\hat{\Lambda}_T^2)^{-1/2} e^{-\hat{P}^2/(2\hat{\Lambda}_T^2)}$.
When $\hat{\Lambda}_T^2 \ll 1$ our beam can be considered cold
meaning that $\hat{F}(\hat{P}) \simeq \delta(\hat{P})$. However
note that, in order to specify quantitatively the range of
validity of the cold beam assumption with respect to the values of
$\hat{\Lambda}_T^2$, one should first solve the more generic
evolution problem for a non-cold beam and then study the limit for
small values of $\hat{\Lambda}_T^2$.

Now Eq. (\ref{integrapr2}) can be expressed in the final,
dimensionless form:

\begin{eqnarray}
\hat{j}_1 = \int_{-\infty}^{\infty} d{\hat{P}}
\left(\hat{a}_{1d}\hat{F}+\hat{a}_{1e}
{d\hat{F}\over{d\hat{P}}}\right) e^{- i {\hat{P}}{\hat{z}}} && \cr
- S_0 \int_0^{\hat{z}} d\hat{z}' \left[\hat{E}_z
\int_{-\infty}^{\infty} d\hat{P} {d \hat{F}\over{d \hat{P}}} e^{i
{\hat{P}(\hat{z}'-\hat{z})}} \right]~, \label{integrapr3}
\end{eqnarray}
where $\hat{a}_{1d}=- e c a_{1d}/J_0$ and $\hat{a}_{1e}=- e c
e^{i\psi_0} a_{1e}/(J_0 \rho \mathcal{E}_0)$. One can combine Eq.
(\ref{integrapr3}) and Eq. (\ref{max4}) in order to obtain a
single integrodifferential equation for $\hat{E}_z$ or,
alternatively, an integral equation for $\hat{j}_1$.

As regards the description of the evolution in terms of
$\hat{E}_z$, direct substitution of Eq. (\ref{integrapr3}) in Eq.
(\ref{max4}) yields immediately

\begin{eqnarray}
\hat{\nabla}_\bot^2 \hat{E}_z  - q^2 \hat{E}_z = i q^2
\int_{-\infty}^{\infty} d{\hat{P}} \left(\hat{a}_{1d}\hat{F}
+~\hat{a}_{1e} {d\hat{F}\over{d\hat{P}}}\right)  e^{- i
{\hat{P}}{\hat{z}}} &&\cr - i q^2 S_0 \int_0^{\hat{z}} d\hat{z}'
\left[\hat{E}_z \int_{-\infty}^{\infty} d\hat{P} {d \hat{F}\over{d
\hat{P}}} e^{i {\hat{P}(\hat{z}'-\hat{z})}} \right]~.&&\cr
\label{integrodiff}
\end{eqnarray}
On the other hand, the description of our system in terms of
$\hat{j}_1$ can be obtained first by solving Eq. (\ref{max4}) and
then substituting the solution in Eq. (\ref{integrapr2}). For the
solution of Eq. (\ref{max4}) we can use the following result (see
for example \cite{FELT}):

\begin{equation}
\hat{E}_z = - {i q^2\over{2\pi}} \int{d{\bf{\hat{r}_\bot^{(s)}}}
\hat{j}_1 K_0\left(q
\mid{\bf{\hat{r}_\bot}}-{\bf{\hat{r}_\bot^{(s)}}}\mid\right)}~,
\label{ez1}
\end{equation}
where $K_0$ indicates the modified Bessel function of the second
kind. Then, substitution in Eq. (\ref{integrapr3}) yields

\begin{eqnarray}
\hat{j}_1 = \int_{-\infty}^{\infty} d{\hat{P}}
\left(\hat{a}_{1d}\hat{F}+\hat{a}_{1e}
{d\hat{F}\over{d\hat{P}}}\right) e^{- i {\hat{P}}{\hat{z}}} &&\cr
+ {i q^2\over{2\pi}} S_0 \int_0^{\hat{z}} d\hat{z}' \left[
\int{d{\bf{\hat{r}_\bot^{(s)}}} \hat{j}_1 K_0\left(q
\mid{\bf{\hat{r}_\bot}}-{\bf{\hat{r}_\bot^{(s)}}}\mid\right)}
\right.&&\cr \left. \times \int_{-\infty}^{\infty} d\hat{P} {d
\hat{F}\over{d \hat{P}}} e^{i {\hat{P}(\hat{z}'-\hat{z})}}
\right]~. \label{integralcurr}
\end{eqnarray}
Note that the description of the system in terms of fields or
currents is completely equivalent. Using one or the other is only
a matter of convenience; it will turn out in the following
Sections that the description in terms of the fields is
particularly suitable for analytical manipulations, while the
description in terms of currents is advisable in case of a
numerical approach.

It is interesting to explore the asymptote of Eq.
(\ref{integrodiff}) for $q \rightarrow \infty$. In this case one
obtains the following simplified equation for the field evolution:

\begin{eqnarray}
\hat{E}_z= - i \int_{-\infty}^{\infty} d{\hat{P}}
\left(\hat{a}_{1d}\hat{F} +~\hat{a}_{1e}
{d\hat{F}\over{d\hat{P}}}\right)  e^{- i {\hat{P}}{\hat{z}}} &&
\cr + i S_0 \int_0^{\hat{z}} d\hat{z}' \left[\hat{E}_z
 \int_{-\infty}^{\infty}
d\hat{P} {d \hat{F}\over{d \hat{P}}} e^{i
{\hat{P}(\hat{z}'-\hat{z})}} \right]~. \label{integro}
\end{eqnarray}
In the case of a cold beam $\hat{F}(\hat{P}) = \delta(\hat{P})$
and Eq. (\ref{integro}) transforms to

\begin{equation}
\hat{E}_z= - i  \left(\hat{a}_{1d}+ i\hat{a}_{1e}\hat{z}\right)
 +  S_0 \int_0^{\hat{z}} d\hat{z}' \left(\hat{z}'-
\hat{z}\right) \hat{E}_z~. \label{integro2}
\end{equation}
Finally, after double differentiation with respect to $\hat{z}$ we
get back the well-known pendulum equation for one-dimensional
systems:

\begin{equation}
{\partial^2 \hat{E}_z\over{\partial \hat{z}^2}}+ S_0 \hat{E}_z =
0~. \label{integro3}
\end{equation}
In this approximation, and in the particular case $S_0 = 1$, we
recover the one-dimensional plasma wave number $\Lambda_p$, which
agrees with the normalization $\hat{z} = \Lambda_p z$. It should
be noted that such a normalization is natural in the limit $q
\rightarrow \infty$ but it progressively loses its physical
meaning as $q$ becomes smaller and smaller: of course, using our
dimensionless equations for $q\ll 1$ will still yield correct
results, because the equations are correct, but the normalization
fits no more the physical feature of the system, in that case.

Eq. (\ref{integro}), which describes the system evolution in the
limit $q \rightarrow \infty$,  corresponds to the the
one-dimensional case. We can use Eq. (\ref{integro}) and impose
$\hat{z}=0$ thus getting the electromagnetic field at the
beginning of the evolution, i.e. the solution of the
electromagnetic problem in the limit $q \rightarrow \infty$:

\begin{equation}
\hat{E}_z = -i \hat{j}_{1\mid_{\hat{z}=0}}~. \label{sole}
\end{equation}
This can be easily written in dimensional form as

\begin{equation}
E_z = {4i\over{k_m r_0^2}} \left( -{I_1\over{c}}\right)~,
\label{soledim}
\end{equation}
where $j_1 \simeq I_1/(\pi r_0^2)$.

We can check Eq. (\ref{soledim}) with already known results in
scientific literature. In fact, the field generated by an electron
beam modulated in density in free space can be easily calculated
in the system rest frame (see e.g. \cite{ROSE} and \cite{EEMF}).
In the limit of a pancake beam (i.e. for large transverse
dimension) the following impedance per unit length of drift has
been found:

\begin{equation}
Z = {4 i\over{k_m r_0^2}}~. \label{imp}
\end{equation}
The first factor on the right hand side of Eq. (\ref{soledim}),
which is the impedance per unit drift according to our
calculations, is exactly the result in Eq. (\ref{imp}) which
proves that our starting equation Eq. (\ref{integrodiff}) can be
used to solve correctly the electromagnetic problem in the case
$q\rightarrow \infty$, as it must be.

Finally, before proceeding, it is worth to estimate the value of
our parameters for some practical example and to see how our
assumptions compare with an interesting case. Nowadays
photo-injected LINACs, to be used as linear colliders or FEL
injectors constitute cutting edge technology as regards electron
particle accelerators. Currents of about $3~k$A with energy such
that $\gamma \sim 10^3$ can be achieved together with quite a
small energy spread $\langle (\Delta \mathcal{E})^2 \rangle^{1/2}
\sim 100~ k$eV, radial dimensions of order $r_0 \sim 100 ~\mu$m
and a normalized emittance $\epsilon_n \sim 1$ mm mrad. In these
conditions, a typical modulation wavelength of about $1 ~\mu$m
will lead to a value of $q \sim 1$, which is well outside the
region of applicability of the one-dimensional theory so that a
generalized theory like the ours must be used. It is interesting
to note that, by means of the definitions of $q$ and $\Lambda_p$,
we can write $\rho = [I/(\gamma I_A)]^{1/2} q^{-1}$: this suggests
that a natural measure for the current is indeed $\gamma I_A$
which is, in practice, the Alfven-Lawson current (see \cite{ALFV},
\cite{LAWS}, \cite{OLSO}). In our numerical example $q \sim 1$ and
the value of $\rho$ is actually determined by the ratio $I/(\gamma
I_A)\sim 2\cdot 10^{-4}$. As a result $\rho \sim 10^{-2}$: this
means that our simplified Maxwell equation, Eq. (\ref{max3}), is
valid up to an accuracy of $10^{-2}$. Moreover, in this case,
$\hat{\Lambda}_T^2\sim 10^{-4}$ so that the cold beam case turns
out to be of great practical interest. Finally $\epsilon_n \sim 1$
mm mrad. This corresponds, for $\gamma \sim 10^3$, to a betatron
function $\beta_{f} \sim \gamma r_0^2/\epsilon_n \sim 10$ m. Our
choice of considering the longitudinal motion alone is satisfied
when the period of a plasma oscillation is much shorter than the
period of a betatron oscillation, that is when $\Lambda_p
\beta_{f} \gg 1$. It should be noted, though, that this estimation
cannot have a rigorous mathematical background before a more
comprehensive theory, including the effects of transverse
dynamics, is developed: only then our present theory can be
reduced to an asymptote of a more general situation, and
conditions for its applicability can be derived in a rigorous way.
Keeping this fact in mind, in our example $\Lambda_p \sim 3 \cdot
10^{-1}$m$^{-1}$ which means $\Lambda_p \beta_{f} \sim  3$
signifying that this particular case is at the boundary of the
region of applicability of our theory.

\section{\label{sub:axis} Main result}

Given Eq. (\ref{integrodiff}) with appropriate initial conditions
for $\hat{j}_{1_{\mid_{\hat{z}=0}}}$ it is possible to find an
analytical solution to the evolution problem. The method is
similar to the one used for the solution of the self-consistent
problem in FEL theory (see \cite{DIFF}) and relies on the
introduction of the Laplace transform of $\hat{E}_z$, namely:

\begin{equation}
\bar{E}(p, {\bf{\hat{r}_\bot}}) = \int_0^{\infty} d \hat{z} e^{- p
\hat{z}} \hat{E}_z~, \label{lapl}
\end{equation}
with $Re(p)>0$. The advantage of the Laplace transform technique
is that the evolution equation is transformed from the
integrodifferential equation Eq. (\ref{integrodiff}) into the
following ordinary differential equation:

\begin{equation}
\left[\hat{\nabla}_\bot^2 - q^2(1- i \hat{D} S_0)\right]\bar{E}= i
q^2 \left( \hat{D}_0 \hat{a}_{1d} + \hat{D}\hat{a}_{1e} \right)~,
\label{diff1}
\end{equation}
where
\begin{equation}
\hat{D}_0 = \int_{-\infty}^{\infty} d \hat{P} {\hat{F}\over{p + i
\hat{P}}}~, \label{D0}
\end{equation}
and
\begin{equation}
\hat{D}= \int_{-\infty}^{\infty} d \hat{P} {d \hat{F}/d{\hat
{P}}\over{p + i \hat{P}}} \label{D0D}
\end{equation}
with the boundary conditions $\bar{E} \longrightarrow 0$ for
$|{\bf{\hat{r}_\bot}}| \longrightarrow \infty$ and $\partial
\bar{E}/
\partial {\bf{\hat{r}_\bot}} \longrightarrow 0$ for $|{\bf{\hat{r}_\bot}}| \longrightarrow
\infty$. We can rewrite Eq. (\ref{diff1}) as

\begin{equation}
\mathcal{L}\bar{E} = {f}~, \label{diff2}
\end{equation}
having introduced

\begin{equation}
\mathcal{L} = \hat{\nabla}_\bot^2  +
\hat{g}({\bf{\hat{r}_\bot}},p)~, \label{opera}
\end{equation}
\begin{equation}
{f}({\bf{\hat{r}_\bot}}, p) =  i q^2 \left( \hat{D}_0 \hat{a}_{1d}
+ \hat{D}\hat{a}_{1e} \right)~ \label{ff}
\end{equation}
and

\begin{equation}
\hat{g}({\bf{\hat{r}_\bot}}, p) = - q^2(1-i\hat{D} S_0)~.
\label{gg}
\end{equation}
%
Note that only Eq. (\ref{integrodiff}) can benefit from the use of
the Laplace transform but not the integral equation Eq.
(\ref{integralcurr}).

Eq. (\ref{diff2}) is a nonhomogeneous, linear, second-order
differential equation. We are interested in solving Eq.
(\ref{diff2}) for any given $p$ such that $Re(p) > 0$. Solution is
found if we can find the inverse of the operator $\mathcal{L}$,
namely a Green function $\bar{G}$ obeying the given boundary
conditions; in this case we simply have

\begin{equation}
\bar{E} = \int d{\bf{\hat{r}'_\bot}}
{\bar{G}}({\bf{\hat{r}_\bot}},{\bf{\hat{r}'_\bot}})
f({\bf{\hat{r}'_\bot}})~. \label{green1}
\end{equation}
\subsection{Generic approach \label{subsub:deco}}

Depending on the choice of $\hat{g}$, i.e. on the choice of $S_0$,
$\hat{F}$ and $p$, the differential operator $\mathcal{L}$ can
change its character completely making $\mathcal{L}$ more or less
difficult to deal with. For example, the case of a self-adjoint
operator is obviously a simple situation, since its eigenvalues
are real and its eigenfunctions form a complete and orthonormal
set for the space of squared-integrable functions $L^2$ (defined
over the entire transverse plane through ${\bf{\hat{r}_\bot}}$)
with respect to the internal product:

\begin{equation}
<h~|~g> = \int d{\bf{\hat{r}_\bot}} g h^*~. \label{inn}
\end{equation}
Then, such a orthonormal set can be used to provide, quite
naturally, an expansion for $\bar{G}$.

However, in the most general situation, $\mathcal{L}$ is not
self-adjoint: to see this, it is sufficient to note that $\hat{g}$
is not real. As a result, the eigenvalues of $\mathcal{L}$ are not
real, its eigenfunctions are not orthogonal with respect to the
internal product in Eq. (\ref{inn}), we do not know wether the
spectrum of $\mathcal{L}$ is discrete, completeness is not granted
and we cannot prove the existence of a set of eigenfunctions
either.

To the best of our knowledge there is no theoretical mean to
really deal with our problem in full generality. When not
self-adjoint operators are encountered in different branches of
Physics (see, for example, \cite{KRIN} and \cite{SIEG})
mathematical rigorousness is somehow relaxed assuming, rather than
proving, certain properties of the operator. We will do the same
here  assuming, to begin, the existence of eigenfunction sets;
then, as for example has been remarked in \cite{KRIN} and
\cite{SIEG} one can consider, together with the spectrum of
$\mathcal{L}$ defined by the eigenvalue problem:

\begin{equation}
\mathcal{L} \Psi_{j} = \Lambda_j \Psi_{j} \label{spec1}
\end{equation}
also the spectrum of its adjoint, defined by
\begin{equation}
\mathcal{L}^{*} {\Psi}_{j}^{*} = [\Lambda_j]^{*} {\Psi}_{j}^{*}~.
\label{spec2}
\end{equation}
It can be shown by using the bi-orthogonality theorem \cite{BIOR}
that

\begin{equation}
<\Psi_{j}^*|{\Psi}_{i}>=\int d{\bf{\hat{r}_\bot}} \Psi_{j}\Psi_{i}
= \delta_{ji}~. \label{orto}
\end{equation}
In other words the sequence $\{\Psi_{j}\}_j$  admits
$\{{\Psi}_{j}^*\}_j$ as a bi-orthonormal sequence. Then one has to
\textit{assume} completeness and discreteness of the spectrum so
that the following expansion is correct:

\begin{equation}
\bar{G}= \sum_j |\bar{G}{\Psi}_{j}><\Psi_{j}^{*}|~, \label{expG1}
\end{equation}
We ascribe to alternative theoretical approaches and numerical
techniques the assessment of the validity region of this
assumption, which should be ultimately formulated  in terms of a
restriction on the possible choices of $S_0$ and $\hat{F}$. In
other words we give here a general method for solving our problem
which is valid only under the fulfillment of certain assumptions,
but we make clear that it is not possible, to the best of our
knowledge, to strictly formulate the applicability region of this
method in terms of properties of $S_0$ and $\hat{F}$ as it would
be desirable.

With this in mind one can use the fact that $\bar{G} \equiv
\mathcal{L}^{-1}$ and write

\begin{equation}
\bar{G}({\bf{\hat{r}_\bot}},{\bf{\hat{r}'_\bot}})=\sum_j
{\Psi_{j}({\bf{\hat{r}_\bot}})\Psi_{j}({\bf{\hat{r}'_\bot}})\over{\Lambda_j}}~.
\label{expGf}
\end{equation}
Finally, substituting Eq. (\ref{expGf}) in Eq. (\ref{green1}) one
gets

\begin{equation}
\bar{E} = \sum_j {\Psi_{j}({\bf{\hat{r}_\bot}})\over{\Lambda_j}}
\int d{\bf{\hat{r}'_\bot}} \Psi_{j}({\bf{\hat{r}'_\bot}})
f({\bf{\hat{r}'_\bot}})~. \label{greenf}
\end{equation}
To find $\hat{E}_z$ we use the inverse Laplace transformation that
is the Fourier-Mellin integral:

\begin{equation}
\hat{E}_z(\hat{z},{\bf{\hat{r}_\bot}}) = {1\over{2 \pi i}}
\int_{\alpha - i\infty}^{\alpha + i\infty} dp
\bar{E}(p,{\bf{\hat{r}_\bot}}) e^{p\hat{z}}~, \label{invLapl}
\end{equation}
where the integration path in the complex $p$-plane is parallel to
the imaginary axis and the real constant $\alpha$ is positive and
larger than all the real parts of the singularities of $\bar{E}$.

The application of the Fourier-Mellin formula comes with another,
separate mathematical problem related with the ability of
performing the integral in Eq. (\ref{invLapl}). One method to
calculate the integral is to use numerical techniques and
integrate directly over the path defined, on the complex
$p$-plane, by $Re(p) = \alpha$.

Yet, there is some room for application of analytical techniques
left. In fact, under the hypothesis that $\bar{E}$ is also
defined, except for isolated singularities, as an analytical
function on the \textit{left} half complex $p$-plane \textit{and}
on the imaginary axis and under the hypotesis that
$\bar{E}\rightarrow 0$ uniformly faster than $1/|p|^k$ for a
chosen $k>0$ and for $\mathrm{Arg}(p)$ within $[\pi/2,3\pi/2]$ one
could use Jordan lemma and close the integration contour of Eq.
(\ref{invLapl}) by a semicircle at infinity on the left half
complex $p$-plane. An obvious (and well-known) problem is that
$\bar{E}$ is defined only for $Re(p)>0$ according to Eq.
(\ref{lapl}). Yet, \textit{if} the border points at $Re(p)=0$ are
regular points of  $\bar{E}$ (except for isolated singularities)
then one can consider the (unique) analytical continuation of
$\bar{E}$ along the border, from the original domain of
analyticity (i.e. the points $p$ with $Re(p)>0$ except for
isolated singularities) to the entire complex plane (again,
isolated singularity excluded). Then one can still apply Jordan
lemma on the analytic continuation of $\bar{E}$ (provided that it
obeys the other assumption), because the final result is the
integral in Eq. (\ref{invLapl}) which is uniquely defined by the
original function $\bar{E}$ for $Re(p)>0$.

The problem is trivially solved for the case of a cold beam
because $\hat{F}=\delta(\hat{P})$ so that $\hat{D}_0 = 1/p$ and
$\hat{D} = - i/p^2$. Then Eq. (\ref{greenf}) defines indeed an
analytic function in all points of the complex plane with the
exception of $p=0$ and the points such that $\Lambda_j(p)=0$. All
the hypothesis of Jordan lemma are verified and the method can be
applied without any problem.

The situation is completely different in the case of a generic
energy spread function $\hat{F}$. In fact by inspection of Eq.
(\ref{D0}) and Eq. (\ref{D0D}) one is immediately confronted with
the fact that the integrands in $\hat{D}_0$ and $\hat{D}$  are,
usually, singular at all the points of the imaginary axis
$Re(p)=0$ since the integration in $\hat{P}$ is taken from
$-\infty$ to $+\infty$. As a result the points $Re(p)=0$ are not
regular points of $\bar{E}$ and $\bar{E}$ cannot be analytically
continued through the border $Re(p)=0$.

This problem is the same encountered in the treatment of Landau
damping (see \cite{LAND}). Of course one may follow the solution
proposed by Landau and present particular \textit{definitions} of
$\hat{D}_0$ and $\hat{D}$ at $Re(p)=0$ that are

\begin{equation}
\hat{D}_0 = \mathcal(P)\int_{-\infty}^{\infty} d \hat{P}
{\hat{F}\over{p + i \hat{P}}}+\pi\hat{F}(ip)~~~~~Re(p)=0
\label{D0zero}
\end{equation}
and
\begin{equation}
\hat{D}= \mathcal(P)\int_{-\infty}^{\infty} d \hat{P} {d
\hat{F}/d{\hat {P}}\over{p + i \hat{P}}}
+\pi\hat{F}'(ip)~~~~~Re(p)=0~,\label{D0Dzero}
\end{equation}
so that $D_0$ and $D$ are now regular for $Re(p)=0$ and can be
(uniquely) continued at $Re(p)<0$ by

\begin{equation}
\hat{D}_0 = \int_{-\infty}^{\infty} d \hat{P} {\hat{F}\over{p + i
\hat{P}}}+2\pi\hat{F}(ip)~~~~~Re(p)<0 \label{D0mino}
\end{equation}
and
\begin{equation}
\hat{D}= \int_{-\infty}^{\infty} d \hat{P} {d \hat{F}/d{\hat
{P}}\over{p + i \hat{P}}}
+2\pi\hat{F}'(ip)~~~~~~Re(p)<0~.\label{D0Dmino}
\end{equation}
In this way the definition of $\bar{E}$ could be extended, except
for isolated singularities, to an analytical function on the
entire complex $p$-plane and, if $\hat{F}(\hat{P})$ behaves
relatively well, Jordan lemma can be applied without further
problems. 

Yet, we think that the application of Landau's prescription, i.e.
the definition in Eq. (\ref{D0zero}) and Eq. (\ref{D0Dzero})
should be taken with extreme caution. As it is reviewed by
Klimontovich \cite{KLIM} and references therein, Landau's method
is equivalent to the introduction of additional assumptions on the
system, namely the adiabatic switching on of the space-charge
field at $t=-\infty$.

Another method equivalent to Landau's consists, as has been
remarked long ago by Lifshitz \cite{LIFT}, in the introduction of
a small dissipative term into the linearized Vlasov equation which
ceases to be \textit{non}-dissipative from the very beginning. The
Vlasov equation is then solved by Fourier technique and the limit
for a vanishing dissipating term is taken in the final result,
which leads, in the end, to Landau's result. Yet, the limit for a
vanishing dissipation must be taken in the final formulas in order
to recover Landau's coefficient \textit{and not before}.
This means that Landau's method consists in the introduction of
additional assumptions regarding the system under study or
equivalently, in changing the very nature of the equations
describing our system (from non-dissipative to dissipative):
therefore in practical calculations, we prefer to deal only with
the case of a cold beam where the original non-dissipative nature
of the system is maintained without problems, leaving the other
case to future study. Such a viewpoint constitutes a restriction
but as we have seen in the previous Section the cold beam case is
practically quite an important issue: in this Section, we will
present our results in full generality without fixing $\hat{F}$
but keeping in mind, however, all the warnings discussed before.

In any case and again, in all generality, we can say that wether
or not the conditions of Jordan lemma are satisfied depends on the
distribution $\hat{F}(\hat{P})$. In the case Jordan lemma is
applicable one can find a closed, analytic expression for
$\hat{E}_z$:

\begin{eqnarray}
\hat{E}_z(\hat{z},{\bf{\hat{r}_\bot}}) = \sum_j
\Phi_{j}({\bf{\hat{r}_\bot}}) e^{\lambda_j \hat{z}}\left[\left({d
\Lambda_j(p)\over{dp}}\right)_{p=\lambda_j}\right]^{-1}&& \cr
\times \int_0^{\infty}
d{\bf{\hat{r}'_\bot}}\Phi_{n}(\bf{\hat{r}'_\bot})
f({\bf{\hat{r}'_\bot}},\lambda_j)~, \label{final1}
\end{eqnarray}
where
$\Phi_{j}({\bf{\hat{r}_\bot}})=\Psi_{j}({\bf{\hat{r}_\bot}},p=\lambda_j)$
and $\lambda_j$ are solutions of the equations:

\begin{equation}
\Lambda_j(p) = 0~ \label{eigenvalue}
\end{equation}
or, which is the same, solution of Eq. (\ref{spec1}) as
$\Lambda_j=0$: from this viewpoint the functions $\Phi_{nj}$
constitute the kernel of the operator $\mathcal{L}$ and
$\lambda_j$ are the values of $p$ such that $\mathcal{L}$ admits a
non-empty kernel.

It is interesting to note that $\{\Phi_{j}\}_{j}$ is a subset of
$\{\Psi_{j}\}_{j,p}$ naturally suited to expand any function of
physical interest (the field $\hat{E}_z$). In this sense, one may
say that the fields are subjected to constraints given by Maxwell
equations, which are codified through the operator $\mathcal{L}$;
these constraints are implicitly used during the anti-Laplace
transform process, thus selecting only those $\{\Psi_{j}\}_{j,p}$
of physical interest. In this spirit, although $p$ is
mathematically allowed to span over all the complex plane with
$Re(p)>0$, only the particular values for which $p=\lambda_j$ have
physical meaning in the final result.


An explicit expression for $({d \Lambda_j(p)/{dp}})_
{p=\lambda_j}$ can be found. Using Eq. (\ref{orto}) and Eq.
(\ref{spec1}) we can write

\begin{equation}
\Lambda_j = \int d{\bf{\hat{r}_\bot}}  \Psi_{j} \mathcal{L}
\Psi_{j}~. \label{Lambda}
\end{equation}
Then, differentiating Eq. (\ref{Lambda}) we have

\begin{equation}
{d\Lambda_j\over{d p} }= \int d{\bf{\hat{r}_\bot}} {\partial
g\over{\partial p}}\Psi_{j}^2 \label{LambdaD}
\end{equation}
Our final result is therefore written as follows:

\begin{equation}
\hat{E}_z(\hat{z},{\bf{\hat{r}_\bot}}) = \sum_j  u_j
\Phi_{j}({\bf{\hat{r}_\bot}}) e^{\lambda_j \hat{z}}~,
\label{finalend}
\end{equation}
where the coupling factor $u_j$ is given by

\begin{equation}
u_j={\int d{\bf{\hat{r}'_\bot}}
\Phi_{j}({\bf{\hat{r}'_\bot}})f({\bf{\hat{r}'_\bot}},\lambda_j)\over{~~~~\left[\int
d{\bf{\hat{r}'_\bot}} \left({\partial g\over{\partial
p}}\right)\Psi_{j}^2\right]_{p=\lambda_j}}} ~.\label{coupl}
\end{equation}
Eq. (\ref{finalend}) describes the evolution of the system under
the action of self-fields in a generic way, for any bunch
transverse shape $S_0$, for any choice of local energy spread
$\hat{P}$ and any initial condition (under the assumptions
mentioned before). Our solution is indeed an analysis in
(self-reproducing) propagating eigenmodes of the electric field.

We have seen that, due to the fact that $\mathcal{L}$ is in
general not self-adjoint, the modes $\Psi_{j}$ are not orthogonal
in the sense of Eq. (\ref{inn}) nor, as a consequence, $\Phi_{j}$
are. Moreover, even if $\Psi_{j}$ were orthogonal, $\Phi_{j}$ are
chosen among the $\Psi_{j}$ at different values of $p$ so that
orthogonality of $\Phi_{j}$ with respect to Eq. (\ref{inn}) is
also not granted. It is possible, however, to formulate
appropriate initial conditions to obtain a single propagating mode
as a solution of our self-consistent problem. This demonstrates
that single modes have physical meaning besides being mathematical
tools for function decompositions.

Suppose, for example, that we wish to excite a single mode at
\textit{fixed} values of $j$. On the one hand Eq. (\ref{finalend})
is simplified to

\begin{equation}
\hat{E}_z = u_j \Phi_{j} ~e^{\lambda_j \hat{z}} \label{simplend}
\end{equation}
and, differentiating with respect to $z$, one also obtains

\begin{equation}
{\partial \hat{E}_z\over{\partial \hat{z}}} = \lambda_j
\hat{E}_z~. \label{diffsimplend}
\end{equation}
On the other hand, the evolution equation, Eq. (\ref{integrodiff})
at $\hat{z}=0$ reads:

\begin{equation}
\mathcal{O}\hat{E}_{z_{\mid_{\hat{z}=0}}} = i q^2 \hat{a}_{1d}~,
\label{evz0}
\end{equation}
where we introduced

\begin{equation}
\mathcal{O}=\hat{\nabla}_\bot^2-  q^2~. \label{opo}
\end{equation}
The same Eq. (\ref{integrodiff}) differentiated with respect to
$\hat{z}$ and evaluated at $\hat{z}=0$ gives

\begin{equation}
\mathcal{O}\left({\partial \hat{E}_z\over{\partial
\hat{z}}}\right)_{\mid_{\hat{z}=0}} = - q^2 \hat{a}_{1e}~,
\label{evdz0}
\end{equation}
which may be rewritten using Eq. (\ref{diffsimplend}) as:

\begin{equation}
\mathcal{O}\hat{E}_{z_{\mid_{\hat{z}=0}}}  = -{q^2
\over{\lambda_j}}~ \hat{a}_{1e}~. \label{evdz1}
\end{equation}
Finally, if $\hat{a}_{1e} \neq 0$, comparison of Eq. (\ref{evdz1})
and Eq. (\ref{evz0}) gives:

\begin{equation}
{\hat{a}_{1e}\over{\hat{a}_{1d}}}={-i\lambda_j}~. \label{constr}
\end{equation}
Since for plasma oscillations $\lambda_{j}$ has to be imaginary,
Eq. (\ref{constr}) fixes the phase $\psi_0 = l\pi$ (with $l$
integer number). The actual shape of $\hat{a}_{1d}$ (and
$\hat{a}_{1e}$) is obtained, modulus a multiplicative constant, by
substitution of Eq. (\ref{simplend}), calculated at $\hat{z}=0$,
in Eq. (\ref{evz0}) and it is fixed by the following condition:

\begin{equation}
\hat{a}_{1d} = \mathcal{O}(\Phi_{j})~. \label{condi}
\end{equation}
We will now present some remarkable example of how to apply Eq.
(\ref{finalend}) and explore, in particular cases, the
applicability region of our method.

We will start our exploration discussing the situation of an
axis-symmetric beam which is still quite a generic one. Given the
symmetry of the problem we will make use, from now on, of a
cylindrical (normalized) coordinate system $(\hat{r}, \phi,
\hat{z})$, with obvious meaning of symbols.

Since $\hat{j}_1= \hat{j}_1(\hat{r},\hat{z},\phi)$, $\hat{E}_z =
\hat{E}_z(\hat{r},\hat{z},\phi)$ and $f = f(\hat{r},p,\phi)$, it
is convenient to decompose them in azimuthal harmonics according
to

\begin{equation}
\hat{j}_1(\hat{z};\hat{r},\phi) = \sum_{n=-\infty}^{\infty}
\hat{j}_1^{(n)}(z,\hat{r}) e^{-i n \phi}~, \label{decomp}
\end{equation}
\begin{equation}
\hat{E}_z(\hat{z};\hat{r},\phi) = \sum_{n=-\infty}^{\infty}
\hat{E}_z^{(n)}(\hat{z};\hat{r}) e^{-i n \phi}~
\label{decompfield}
\end{equation}
and
\begin{equation}
f(\hat{r},p,\phi) = \sum_{n=-\infty}^{\infty} f^{(n)}(\hat{r},p)
e^{-i n \phi}~. \label{decompf}
\end{equation}

Moreover, in cylindrical coordinates we have

\begin{equation}
\hat{\nabla}^2_\bot = {1\over{\hat{r}}}{\partial \over{\partial
\hat{r}}}\left[\hat{r}{\partial \over{\partial \hat{r}}}\right]+
{1\over{r^2}}{\partial^2\over{\partial \phi^2}}~. \label{nablacyl}
\end{equation}

These definitions allow us to write our equations and results for
the $n$-th azimuthal harmonic of the electric field. In this
situation the operators $\mathcal{L}$ and $\mathcal{O}$ can be
written as

\begin{equation}
\mathcal{L} = {\partial^2 \over{\partial \hat{r}^2}} +
{1\over{\hat{r}}}{\partial \over{\partial \hat{r}}} -
{n^2\over{\hat{r}^2}} + \hat{g}(\hat{r},p)~, \label{operallo}
\end{equation}
where, now, $S_0=S_0(\hat{r})$ in the definition of $\hat{g}$ and

\begin{equation}
\mathcal{O}={\partial^2 \over{\partial \hat{r}^2}} +
{1\over{\hat{r}}}{\partial \over{\partial \hat{r}}} - \left( q^2+
{n^2\over{\hat{r}^2}}\right)~ \label{operopo}
\end{equation}
so that

\begin{equation}
\mathcal{L}\bar{E}^{(n)} = {f}^{(n)}~, \label{diff2sym}
\end{equation}
Having specialized our results to the axis-symmetric case it is
worth to spend some words on the nature of the operator
$\mathcal{L}$ in one particular case. As we have said at the
beginning of this section, the case of a self-adjoint operator is
a particularly blessed one. It is interesting to note that in the
axis-symmetric case, when $S_0$, $\hat{F}$ and $p$ are such that
$\hat{g}$ is real, we deal with a singular Sturm-Liouville problem
as it is shown immediately by multiplying side by side by
$\hat{r}$ Eq. (\ref{diff2sym}). In fact, in this case
$\mathcal{L}$ can be written in the usual form presented in
Sturm-Liouville theory:

\begin{equation}
\mathcal{L} = {\partial \over{\partial
\hat{r}}}\left[\hat{r}{\partial \over{\partial \hat{r}}}\right] -
{n^2\over{\hat{r}}} + \hat{r}\hat{g}(\hat{r},p)~.
\label{operausua}
\end{equation}
In this case given the internal (axisisymmetric) product in $L^2$:

\begin{equation}
<h~|~g> = \int_0^{\infty} d\hat{r}\hat{r} g h^*~, \label{inn2}
\end{equation}
self-adjointness condition for $\mathcal{L}$ is satisfied in the
interval $[0,\infty)$ by all the functions in the linear space
$\mathbb{S}$ which we define as the space of integrable-square
functions not singular with their first derivatives at $\hat{r}=0$
and such that, for any $f,g$ chosen in $\mathbb{S}$ the following
condition is satisfied:

\begin{equation}
\lim_{\hat{r} \rightarrow \infty}
\hat{r}\left[g^*(\hat{r})f'(\hat{r})-f(\hat{r})g'^*(\hat{r})\right]=0~.
\label{conadj}
\end{equation}
Note that $\mathbb{S}$ is a subset of $L^2$. 

This is of course a very particular situation, interesting to
discuss but unfortunately not very useful in practise, since in
order to solve our problem we still have to \textit{assume} that
Eq. (\ref{expG1}) is correct for a generic $p$. When this
assumption is made, in the axis-symmetric case our results Eq.
(\ref{finalend}) and Eq. (\ref{coupl}) take the form

\begin{equation}
\hat{E}_z^{(n)}(\hat{z},\hat{r}) = \sum_j  u_{nj}
\Phi_{nj}(\hat{r}) e^{\lambda^{(n)}_j \hat{z}}~,
\label{finalendsym}
\end{equation}
where

\begin{equation}
u_{nj}(\hat{r})={\int_0^{\infty} d\hat{r}'\hat{r}'\Phi_{nj}f^{(n)}
(\hat{r}',\lambda_j^{(n)})\over{~~~~\left[\int_0^{\infty}
d\hat{r}'\hat{r}' \left({\partial g\over{\partial
p}}\right)\Psi_{nj}^2\right]_{p=\lambda^{(n)}_j}}}
~.\label{couplsym}
\end{equation}
Within this special situation we will now treat in detail the case
of a stepped or parabolic transverse profile. Further on we will
see how the solution for the stepped transverse profile can be
used to obtain a semi-analytic solution for any transverse
profile.

\subsection{\label{subsub:step} Stepped profile}

Consider the case of a step function $S_0 = 1$ for $\hat{r}<1$ and
$S_0=0$ for $\hat{r}\geq 1$. In this case $\hat{g}(\hat{r},p)$
inside the operator $\mathcal{L}$ is simply given by

\begin{equation}
\hat{g}(\hat{r}, p) =\left\{
\begin{array}{c}
- q^2\left(1+{1\over{p^2}}\right) ~~~~\hat{r}<1\\
- q^2                             ~~~~~~~~~~~~~~~~~\hat{r}\geq 1
\end{array}
\right.~,
\label{simplG}
\end{equation}
Let us restrict to the assumption of a cold beam with $F(\hat{P})=
\delta(\hat{P})$. Then $\hat{D}= i/p^2$.

First we look for the solutions of Eq. (\ref{spec1}) with the
boundary condition that $\psi_{nj}$ and their first derivatives
vanish at infinity. The search for the eigenfunctions can be
broken down into an internal ($\hat{r}<1$) and an external
($\hat{r}>1$) problem, with the conditions of continuity for
$\psi_{nj}$ and its derivative across the boundary, since the
final result, the electric field, is endowed with these properties
too. We recognize immediately that the internal and the external
problems are, respectively, the complex Bessel and modified Bessel
equations with appropriate boundary conditions.

Keeping in mind the physical nature of our problem, we impose that
$\psi_{nj}$ must be regular functions of $\hat{r}$ over
$[0,\infty)$. Then, without loss of generality, we can exclude
Bessel functions $Y_n$ and $I_n$ from entering our expression for
$\psi_{nj}$.

Since for the field calculations we are interested in finding the
eigenfunctions $\phi_{nj}= \psi_{nj}(\hat{r},p=\lambda^{(n)}_j)$
we can impose $\Lambda_j^{(n)}=0$, thus obtaining

\begin{equation}
\phi_{nj}(\hat{r}) =\left\{
\begin{array}{c}
C_1 J_n(\alpha_j \hat{r})~~~~~~~~~~\hat{r}<1\\
C_2 K_n(q \hat{r}) ~~~~~~~~~~\hat{r}\geq 1
\end{array}
\right.~, \label{solpsi}
\end{equation}
where $\alpha_j \equiv \sqrt{\hat{g} (\hat{r},p)_{\mid_{
p=\lambda_j^{(n)},\hat{r}<1}}}$ and $\lambda_j^{(n)}$ are roots of
$\Lambda_j^{(n)}(p)=0$, to be still determined at this stage.
Imposing continuity of $\psi_{nj}$ and its derivative at $\hat{r}
= 1$ one finds

\begin{equation}
C_2 = C_1  {J_n(\alpha_j)\over{K_n(q)}}~,\label{conpsi1}
\end{equation}
which leaves the choice of an unessential multiplicative constant,
and

\begin{eqnarray}
\alpha_j J'_{n}(\alpha_j) K_n(q)- q K'_{n}(q) J_n(\alpha_j) = 0~.
\label{conpsi2}
\end{eqnarray}
Eq. (\ref{conpsi2}) can be rewritten with the help of recurrence
relations for Bessel functions in the following form:

\begin{equation}
\alpha_j J_{n+1}(\alpha_j)K_n(q)  - q K_{n+1}(q) J_n(\alpha_j) = 0
\label{conpsi2bis}~.
\end{equation}
Eq. (\ref{conpsi2bis}) is our eigenvalue equation, defining the
values of $\alpha_j$ or, equivalently, of $\lambda^{(n)}_j$. Since
$q$ is real and positive one must have that $\alpha_j
J_{n+1}(\alpha_j)/J_n(\alpha_j)$ is real and positive. Then, it
can be shown that $\alpha_j$ must be real. As a result
$\lambda_j^{(n)}$ are imaginary and such that
$-1<Im(\lambda_j^{(n)})<1$. For any given eigenvalue
$\lambda_j^{(n)}$, also $-\lambda_j^{(n)}$ is solution of Eq.
(\ref{conpsi2bis}) corresponding to fast and slow plasma waves:
from now on we will consider, for simplicity of notation, only the
branch $Im(\lambda_j^{(n)})>0$. Note that from a physical
viewpoint, the condition that $\lambda_j^{(n)}$ is imaginary means
that we are in the absence of damped or amplified oscillations. On
the other hand, the fact that $Im(\lambda_j^{(n)})<1$ means that
plasma oscillations have a minimum wavelength given by $2 \pi
/\Lambda_p$.

\begin{figure}
\begin{center}
\includegraphics[width=90mm]{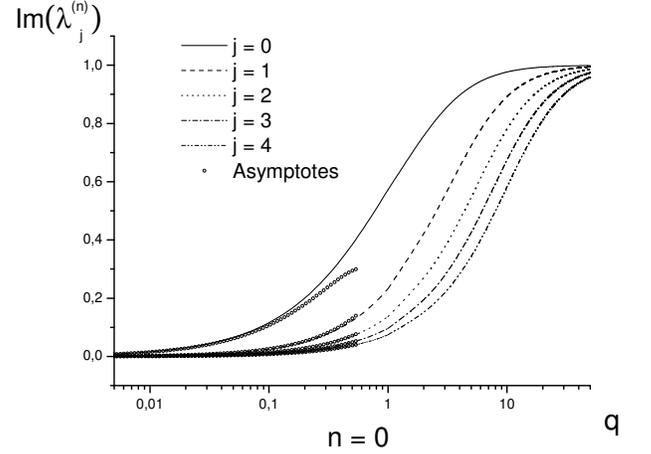}
\caption{\label{eig0} The first five (imaginary) eigenvalues
$\lambda^{(n)}_j$ in units of $i$ as a function of $q$ for $n=0$.
}
\end{center}
\end{figure}

\begin{figure}
\begin{center}
\includegraphics[width=90mm]{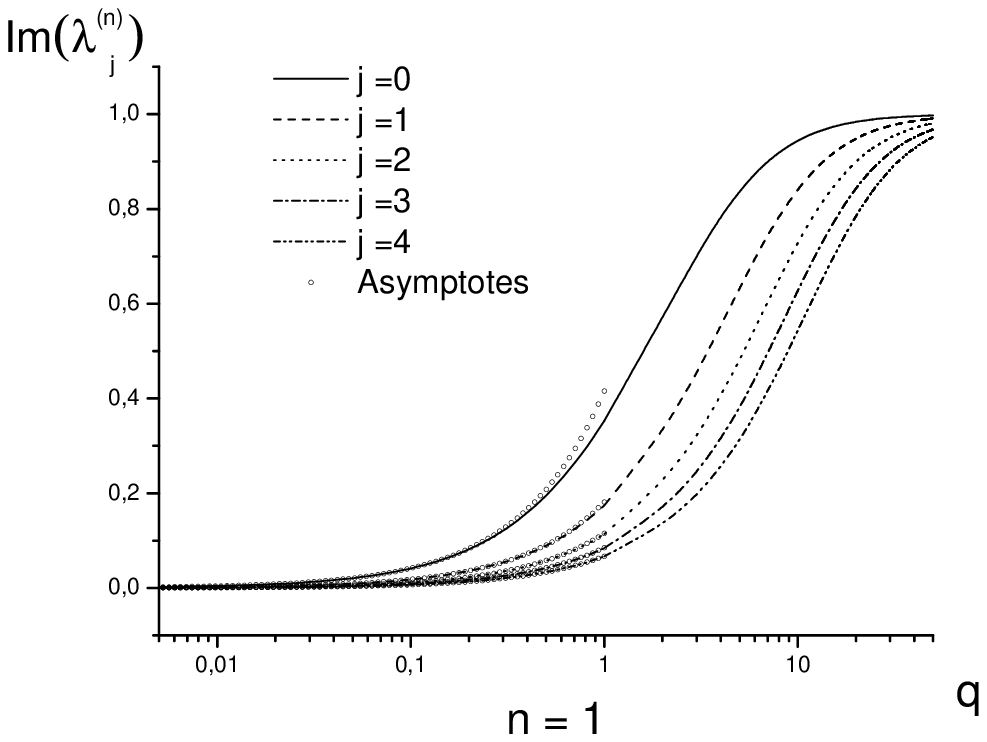}
\caption{\label{eig1} The first five (imaginary) eigenvalues
$\lambda^{(n)}_j$ in units of $i$ as a function of $q$ for $n=1$.
}
\end{center}
\end{figure}

\begin{figure}
\begin{center}
\includegraphics[width=90mm]{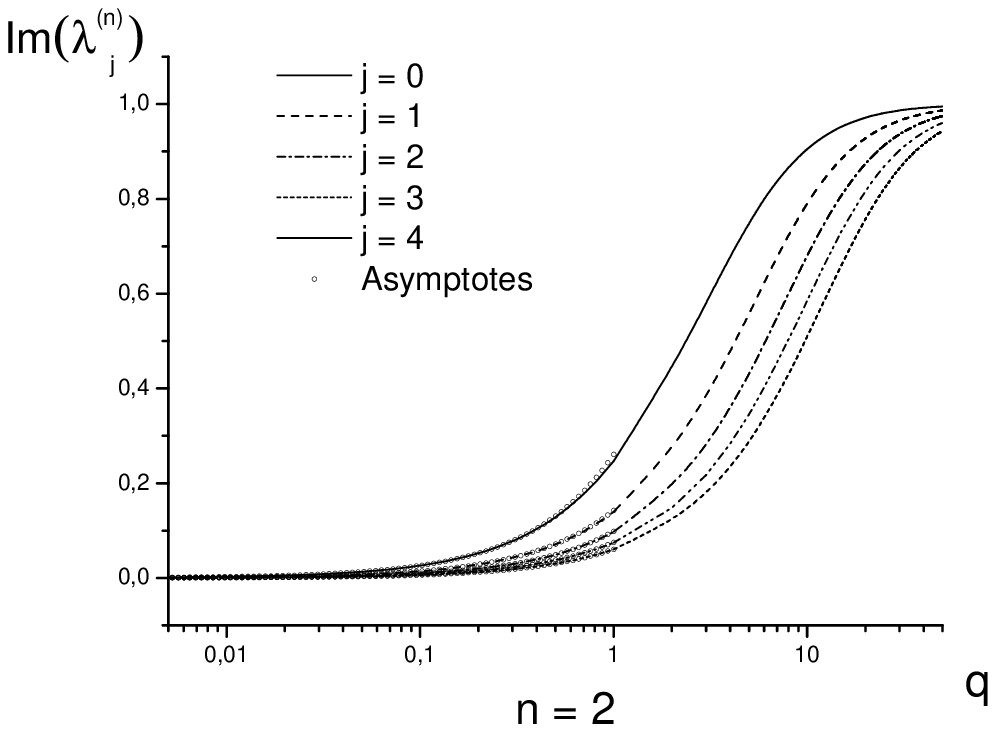}
\caption{\label{eig2} The first five (imaginary) eigenvalues
$\lambda^{(n)}_j$ in units of $i$ as a function of $q$ for $n=2$.
}
\end{center}
\end{figure}

It is interesting to plot the behavior of $Im(\lambda^{(n)}_j)$,
parameterized for several values of $n$ and $j$ as a function of
$q$. Fig. \ref{eig0}, Fig. \ref{eig1} and Fig. \ref{eig2} show the
behavior, as a function of $q$, of the first five eigenvalues for
the first three azimuthal harmonics. It should be noted that
$Im(\lambda^{(n)}_j)$ increases with $q$ and therefore with $r_0$,
but $\Lambda_p$ scales as $r_0^{-1}$; as a result, the period of
the self-reproducing solution identified by fixed values of $n$
and $j$, that is $2 \pi/(\Lambda_p Im(\lambda^{(n)}_j))$, will
increase as $r_0$ is increased. As it can be seen by inspection
all the imaginary parts of the eigenvalues converge to $1$ as
$q\rightarrow \infty$; this can also be derived directly from Eq.
(\ref{conpsi2bis}). As $q\rightarrow \infty$ we have $K_n(q)\simeq
K_{n+1}(q)$, so that Eq. (\ref{conpsi2bis}) gives simply
$J_n(\alpha_j) =0$; this is possible only when $\alpha_j =
\nu_{n,j}$, where $\nu_{n,j}$ is the $j$-th root of $J_n$. Then,
in this limit, $\left(\lambda_j^{(n)}\right)^2 =
-q^2/(q^2+\nu_{n,j}^2)$ and
$\left(\lambda_j^{(n)}\right)^2\rightarrow -1$ since $q\rightarrow
\infty$. Note that convergence to unity tends to get slower as $n$
and $j$ increase. On the other hand, when $q$ becomes smaller and
smaller the plasma wavelength associated with each mode starts to
differ significantly from $\Lambda_p$ and, as noted before, we
should use an effective $\tilde{\Lambda}_p$ in place of
$\Lambda_p$ in our dimensionless quantities in order for these to
retain their physical insight.

The asymptotic behavior of $Im(\lambda)$ for $q \ll 1$ can be
derived directly from Eq. (\ref{conpsi2bis}) too. We consider
first the case $n=0$. In the limit $q \ll 1$ we have $q
K_1(q)/K_0(q) \sim -(\ln q - \ln 2 + \gamma_E)^{-1}$, where
$\gamma_E$ is the Euler gamma constant. We remember that $x
J_1(x)/J_0 \sim x^2/2$ for $x^2 \ll 1$ and that
$\left(\lambda_j^{(n)}\right)^2 \sim -q^2/(\alpha_{j}^2)$.
Neglecting $- \ln 2 + \gamma_E$ we easily find
$\left(\lambda_0^{(0)}\right)^2 \sim q^2 \ln q$. This result is
only valid for $q^2/\lambda^2 \ll 1$ which corresponds, once
plotted in Fig. \ref{eig0}, to the solution for $j=0$ only. The
case $j>0$ is solved using the fact that $-(\ln q)^{-1} \ll 1$:
then the eigenvalue equation is solved only when $\alpha_j \sim
\nu_{1,j}$ which means $\left(\lambda_j^{(0)}\right)^2 \sim -
q^2/\nu_{1,j}$ for $j > 0$.

For $n \ne 0$ instead, when $q \ll 1$ we have $q K_1(q)/K_0(q)
\sim 2 n$. Then, since $2 n J_n(x) \sim x J_{n-1}(x) + x
J_{n+1}(x)$, we find $J_{n-1}(\alpha_{j})=0$ and, therefore,
$\left(\lambda_j^{(0)}\right)^2 \sim - q^2/\nu_{n-1,j}$. These
asymptotic limits are compared with the actual solutions of the
eigenvalue equation in Fig. \ref{eig0}, Fig. \ref{eig1} and Fig.
\ref{eig2}.

Note that in the region $q \ll 1$, if the alternative
normalization using $\tilde{\Lambda}_p$ is selected, $\rho$ shows
only a weak logarithmic dependence on the transverse beam size $q$
in the case $n=0$, $j=0$ and no dependence on $q$ in the other
cases. Looking at the slopes in the figures we can conclude that,
with the use of $\tilde{\Lambda}_p$ in place of $\Lambda_p$,
$\rho$ is, for realistic choices of $I$, much smaller than unity.
The same applies when $q\rightarrow \infty$: in this case
$\tilde{\Lambda}_p \simeq \Lambda_p$ and $\rho$ will also be small
with respect to unity. As a result $\rho$, defined using
$\tilde{\Lambda}_p$, can be considered much smaller than unity in
a wide range of parameters which justifies, at least in this
particular situation, the assumptions used in the derivation of
Eq. (\ref{max3}).

We can now write our final solution in the following form:

\begin{equation}
\hat{E}^{(n)}_z(\hat{z},\hat{r})=\left\{
\begin{array}{c}
\sum_{j} u_{nj} J_n(\alpha_j\hat{r}) e^{\lambda_j^{(n)} \hat{z}}
~~~~~~~~~~~~\hat{r}<1\\
\sum_{j} u_{nj} {J_n(\alpha_j)\over{K_n(q)}}K_n(q\hat{r})
e^{\lambda_j^{(n)} \hat{z}} ~~~~~\hat{r}\geq 1
\end{array}
\right.~,
\label{stepfinal}
\end{equation}
and the coupling factors $u_{nj}$ are given by:

\begin{equation}
u_{nj} = {K_n(q)\int_0^1 d\xi J_n(\alpha_j
\xi)\xi{f}^{(n)}(\xi)\over{J_n(\alpha_j) {d\over{dp}}\left[\alpha
J_{n+1}(\alpha)K_n(q)-q K_{n+1}(q)J_n({\alpha})
\right]_{p=\lambda_j^{(n)}}}}~, \label{ujstep}
\end{equation}
where $\alpha\equiv \sqrt{g(\hat{r},p)_{\mid_{\hat{r}<1}}}$.

At this point we should show that the expansion Eq. (\ref{expG1})
is correct, so that the method used up to now can be rightfully
applied. Yet we cannot do this. We assume this fact, instead, and
we prove that this assumption is right both using numerical
techniques in Section \ref{sub:test} and with the help of an
alternative analytical technique here: in fact, interestingly
enough, one can solve Eq. (\ref{diff1}) also by finding directly a
Green function without any particular expansion simply imposing
that the Green function obeys $\mathcal{L} \bar{G}^{(n)} = 0$ for
all $\hat{r}$ except $\hat{r}=\hat{r}'$, where it must be
continuous and such that its derivative is discontinuous (the
difference of the left and right limit must equal $1/\hat{r}$).
Moreover it must be finite at $\hat{r}=0$. We do not work out
details, which can be found in \cite{FELT}, but we underline the
fact that the anti-Laplace transform of $\bar{E}^{(n)}$ calculated
with this method coincides with our previous result Eq.
(\ref{stepfinal}). Note that in the case the Green function is
derived without the expansion in $\Psi_{nj}$ the final solution
for $\hat{E}_z$ is automatically valid, but still subjected to the
assumptions on the validity of Jordan lemma; without the
introduction of other assumptions on the system under study we can
safely say that our result holds for the case of a cold beam only.

\subsection{\label{subsub:para} Parabolic profile}

A parabolic transverse profile corresponds to the case
$S_0(\hat{r}) =1- k_1^2 \hat{r}^2$. This is one of the few
profiles for which the evolution problem can be solved
analytically. The study of this situation offers, therefore, the
possibility of crosschecking analytical and numerical results with
or without the use of the semi-analytical method described in
Section \ref{subsub:mult}.

In the parabolic case $\hat{g}(\hat{r},p)$ inside the operator
$\mathcal{L}$ is given by

\begin{equation}
\hat{g}(\hat{r}, p) =\left\{
\begin{array}{c}
- q^2\left(1+{1-k_1^2\hat{r}^2\over{p^2}}\right) ~~~~\hat{r}<1\\
- q^2       ~~~~~~~~~~~~~~~~~~~~~~~\hat{r}\geq 1
\end{array}
\right.~, \label{simplP}
\end{equation}
Here we assume, strictly, $F(\hat{P})= \delta(\hat{P})$. Solution
for the homogeneous problem defined by $\mathcal{L}$ can be found
in \cite{FELT}, since it is of relevance in FEL theory as well. We
can use that solution in order to solve our eigenvalue problem,
and to write the expressions for the eigenfunctions $\Psi_{nj}$ to
be inserted in Eq. (\ref{finalend}). Let us introduce the
following notations: $\mu^2 = i \hat{D} q^2 - \Lambda^{(n)}_j$,
$\delta^2 = i \hat{D} K_1^2$, $d^2 = \Lambda^{(n)}_j$, $\epsilon =
(n+1)/2 - \mu^2/(4 \delta)$. After some calculation we find:

\begin{equation}
\Psi_{nj}(\hat{r})=\left\{
\begin{array}{c}
\hat{r}^{n} e^{-\delta \hat{r}^2/2} ~_1F_1(\epsilon,n+1,\delta
\hat{r}^2)
~~~~~~\hat{r}<1\\
e^{-\delta/2} ~_1F_1(\epsilon,n+1,\delta) {K_n(d
\hat{r})\over{K_n{d}}} ~~~~~~\hat{r}\geq 1
\end{array}
\right.~. \label{parfinal}
\end{equation}
where$~_1F_1$ is the confluent hypergeometric function, and the
eigenvalue equation analogous of Eq. (\ref{conpsi2bis}) is now

\begin{eqnarray}
\delta K_n(d)\left[2\epsilon(n+1)^{-1} ~_1F_1(\epsilon+1,n+2,
\delta) \right. && \cr \left. - ~_1F_1(\epsilon,n+1,
\delta)\right] + d K_{n+1}(d) ~_1F_1(\epsilon,n+1, \delta) =
0\label{eigepara}
\end{eqnarray}

Once more we should show that the expansion Eq. (\ref{expG1}) is
correct. We assume this fact, and we present a cross-check of our
result Eq. (\ref{parfinal}) using numerical techniques in Section
\ref{sub:test}.

\subsection{\label{subsub:mult} Multilayer method approach}

An arbitrary gradient axisymmetric profile can be approximated by
means of a given number of stepped profiles, or layers,
superimposed one to the other. This means that results in Section
\ref{subsub:step} can be used to construct an algorithm to deal
with any profile (see \cite{FELT} or \cite{DIFF}  for more details
and a comparison with the same technique in FEL physics).

The normalized radius of the beam boundary is simply unity; let us
divide the region $0<\hat{r}<1$ into $K$ layers, assuming that the
beam current is constant within each layer. Within each layer $k$,
according to Eq. (\ref{integrodiff}), the solution for the
eigenfunction is of the form

\begin{equation}
\Phi_n^{(k)} = c_k J_n(\mu_k \hat{r}) + d_k N_n(\mu_k \hat{r})~,
\label{sollay}
\end{equation}
where $(k-1)K<\hat{r}<k/K$, $c_k$ and $d_k$ are constants, $J_n$
and $N_n$ are the Bessel functions of first and second kind of
order $n$, and

\begin{equation}
\mu_k^2 =- q^2(1-i\hat{D} S_{k-1/2})~.
\label{mu}
\end{equation}
Here $S_{k-1/2} = S_0(\hat{r}_{k-1/2})$ and
$\hat{r}_{k-1/2}=(k-1/2)/K$. To avoid singularity of the
eigenfunction at $\hat{r} = 0$ we should let $d_1 =0$. Then, the
continuity conditions for the eigenfunctions and its derivative at
the boundaries between the layers allow one to find all the other
coefficients. The continuity conditions can be expressed in matrix
form in the following way:

\begin{equation}
\left(
\begin{array}{c}
c_{k+1}\\ d_{k+1}
\end{array}
\right) = T_k
\left(
\begin{array}{c}
c_{k}\\ d_{k}
\end{array}
\right)~, k=1,2,...,K-1~,
\label{matrixco}
\end{equation}
where the coefficients $T_k$ are given by $(\hat{r}_k= k/K)$:

\begin{eqnarray}
(T_k)_{11} ~=~ (\pi/2)\hat{r}_k\left[\mu_k
J_{n+1}(\mu_k\hat{r}_k)N_n(\mu_{k+1}\hat{r}_k)\right.&&\cr\left.
-\mu_{k+1}J_n(\mu_k\hat{r}_k)N_{n+1}(\mu_{k+1}\hat{r}_k)\right]~,&&
\cr (T_k)_{12} ~=~ (\pi/2)\hat{r}_k\left[\mu_k
N_{n+1}(\mu_k\hat{r}_k)N_n(\mu_{k+1}\hat{r}_k)\right.&&\cr\left.
-\mu_{k+1}N_n(\mu_k\hat{r}_k)N_{n+1}(\mu_{k+1}\hat{r}_k)\right]~,&&
\cr (T_k)_{21} ~=-(\pi/2)\hat{r}_k\left[\mu_k
J_{n+1}(\mu_k\hat{r}_k)J_n(\mu_{k+1}\hat{r}_k)\right.&&\cr\left.
-\mu_{k+1}J_n(\mu_k\hat{r}_k)J_{n+1}(\mu_{k+1}\hat{r}_k)\right]~,&&
\cr (T_k)_{22} ~=-(\pi/2)\hat{r}_k\left[\mu_k
N_{n+1}(\mu_k\hat{r}_k)J_n(\mu_{k+1}\hat{r}_k)\right.&&\cr\left.
-\mu_{k+1}N_n(\mu_k\hat{r}_k)J_{n+1}(\mu_{k+1}\hat{r}_k)\right]~.&&
\cr \label{matrelem}
\end{eqnarray}
Eq. (\ref{integrodiff}) also dictates the form of the solution for
the eigenfunction outside the beam $\hat{r}\geq 1$, satisfying the
condition of quadratic integrability:

\begin{equation}
\Phi_{n}(\hat{r}) = b K_n(q\hat{r}),~Re(q)>0~. \label{outsol}
\end{equation}
Then, continuity at the boundary, i.e. at $\hat{r} = 1$ gives the
following relations:

\begin{equation}
c_k J_n(\mu_k) + d_k N_n(\mu_k) = b K_n(q) \label{bound}
\end{equation}
and

\begin{equation}
\mu_k c_k J_{n+1}(\mu_k) + \mu_k d_k N_{n+1}(\mu_k) = b
K_{n+1}(q)~. \label{bound2}
\end{equation}
The two relations above can be also written in matrix form as:

\begin{equation}
T_K\left(
\begin{array}{c}
c_{k}\\ d_{k}
\end{array}
\right) = b
\left(
\begin{array}{c}
1 \\ 1
\end{array}
\right)~,
\label{matbound}
\end{equation}
where the coefficient $b$ can be expressed in terms of the
coefficient $c_1$ by multiple use of Eq. (\ref{matrixco}). Since
$c_1$ can be chosen arbitrarily, we may set $c_1=1$ to obtain the
following matrix equation:

\begin{equation}
T_K \times T_{K-1} \times ... \times T_1\left(
\begin{array}{c}
1 \\ 0
\end{array}
\right) = T\left(
\begin{array}{c}
1 \\ 0
\end{array}
\right) = b\left(
\begin{array}{c}
1 \\ 1
\end{array}
\right)~.
\label{matmid}
\end{equation}
The matrix $T$ depends on the unknown quantity $\hat{\Lambda}$.
The other unknown quantity in Eq. (\ref{matmid}), the coefficient
$b$, can be easily excluded, thus giving the eigenvalue equation:

\begin{equation}
(T)_{11}=(T)_{21}~,
\label{eigenmat}
\end{equation}
which allows one to find the eigenvalue $\hat{\Lambda}$. Then
using Eq. (\ref{matrixco}) and Eq. (\ref{matbound}) it is possible
to calculate the eigenfunction.

\section{\label{sub:test} Applications and exemplifications}

\subsection{\label{subsub:prog} Algorithm for numerical solution}

The results in Section \ref{sub:axis} constitute one of the few
existing solutions for the evolution problem of a system of
particles and field. Yet, to derive it, we had to rely on several
assumptions, among which that of a small perturbation, in order to
get linearized Maxwell-Vlasov equations. This is not too
restrictive, since in practice one has often to deal with
space-charge waves in the linear regime, but it would be
interesting to provide a solution for the full problem. From this
viewpoint, the only way to proceed is the development of some
numerical code based on macroparticle approach capable to deal
with the most generic problem. As a first, initial step towards
this more ambitious goal we present here a numerical solution of
the evolution equation in the case of an axis-symmetric beam, that
we will cross-check with our main result, Eq. (\ref{finalend}). In
order to build a numerical solution one may, in principle, use Eq.
(\ref{integrodiff}), but it turns out more convenient to make use
of Eq. (\ref{integralcurr}).

\begin{figure}
\begin{center}
\includegraphics[width=90mm]{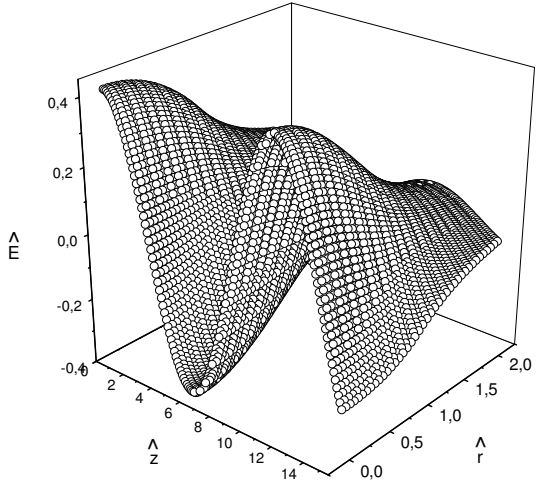}
\caption{\label{Step3D} $\hat{E}=Re(\hat{E}_z)$ as a function of
$\hat{z}$ and $\hat{r}$. Here $q = 1$, $n=0$ and the first five
eigenfunctions have been used. A stepped transverse profile has
been used.}
\end{center}
\end{figure}
\begin{figure}
\begin{center}
\includegraphics[width=90mm]{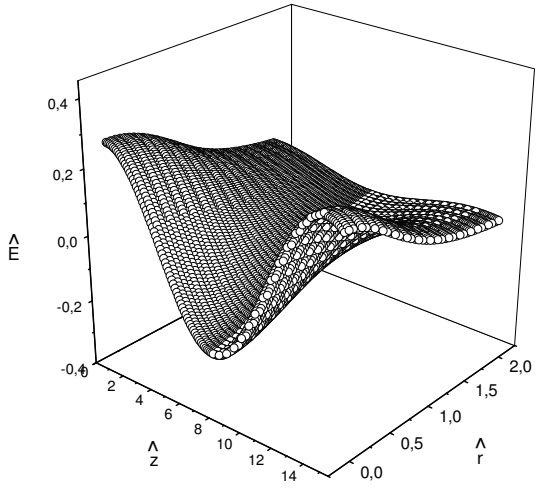}
\caption{\label{Para3D} $\hat{E}=Re(\hat{E}_z)$ as a function of
$\hat{z}$ and $\hat{r}$. Here $q = 1$, $n=0$ and the first five
eigenfunctions have been used. A parabolic transverse profile
proportional to $1-k_1^2 \hat{r}^2$ with $k_1=1.0$ has been used.}
\end{center}
\end{figure}
\begin{figure}
\begin{center}
\includegraphics[width=90mm]{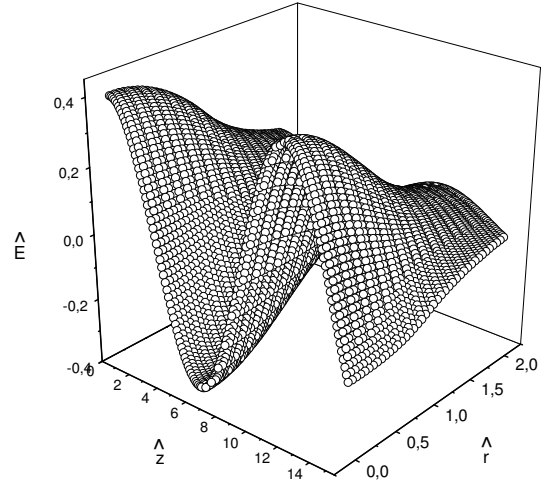}
\caption{\label{Gauss3D} $\hat{E}=Re(\hat{E}_z)$ as a function of
$\hat{z}$ and $\hat{r}$. Here $q = 1$, $n=0$ and the first five
eigenfunctions have been used. A gaussian transverse profile
proportional to $e^{-\hat{r}^2/(2\sigma^2)}$ with $\sigma=2.0$ has
been used.}
\end{center}
\end{figure}
\begin{figure}
\begin{center}
\includegraphics[width=90mm]{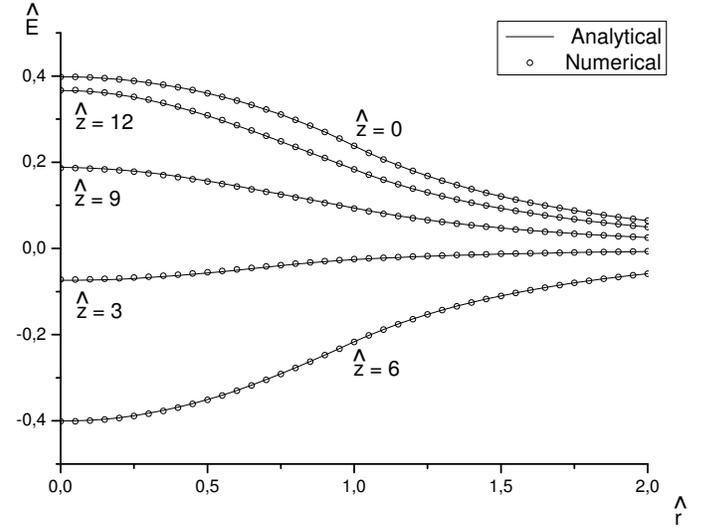}
\caption{\label{cmpS} Comparison between analytical results (solid
line) and numerical methods (circles).  $\hat{E}=Re(\hat{E}_z)$ is
plotted as a function of $\hat{r}$ for several values of
$\hat{z}$. Here $q = 1$, $n=0$ and the first five eigenfunctions
have been used. A stepped transverse profile has been used.}
\end{center}
\end{figure}
\begin{figure}
\begin{center}
\includegraphics[width=90mm]{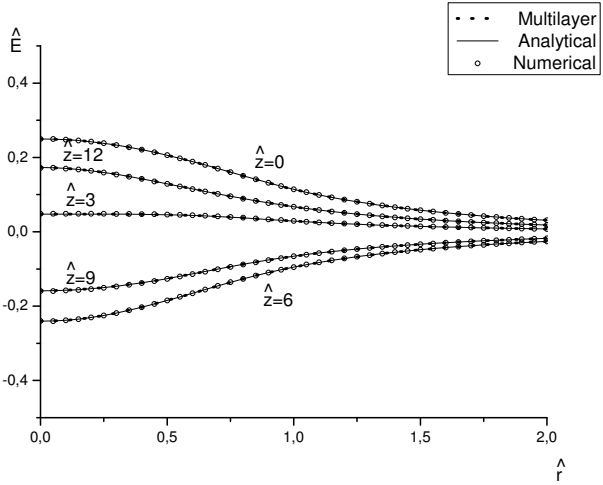}
\caption{\label{cmpP} Comparison between analytical results (solid
line), multilayer approximation with 15 layers (dotted line) and
numerical methods (circles).  $\hat{E}=Re(\hat{E}_z)$ is plotted
as a function of $\hat{r}$ for several values of $\hat{z}$. Here
$q = 1$, $n=0$ and the first five eigenfunctions have been used. A
parabolic transverse profile proportional to $1-k_1^2 \hat{r}^2$
with $k_1=1.0$ has been used.}
\end{center}
\end{figure}
\begin{figure}
\begin{center}
\includegraphics[width=90mm]{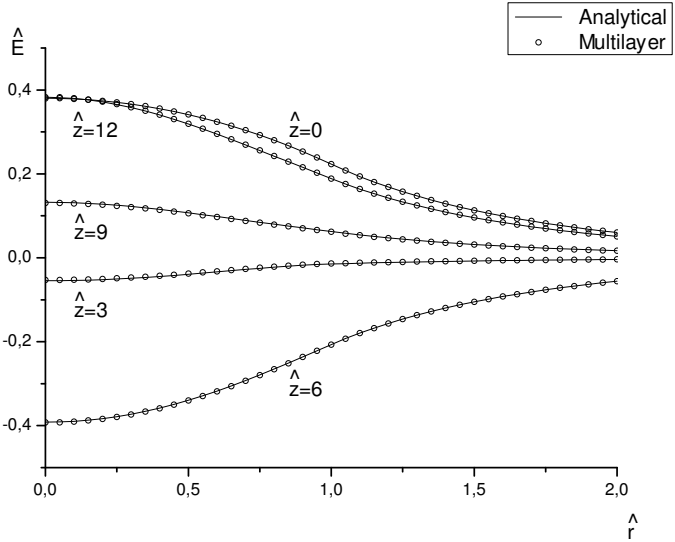}
\caption{\label{cmpG} Comparison between multilayer approximation
with 15 layers (solid line) and numerical  methods (circles).
$\hat{E}=Re(\hat{E}_z)$ is plotted as a function of $\hat{r}$ for
several values of $\hat{z}$. Here $q = 1$, $n=0$ and the first
five eigenfunctions have been used. A gaussian transverse profile
proportional to $e^{-\hat{r}^2/(2\sigma^2)}$ with $\sigma=2.0$ has
been used.}
\end{center}
\end{figure}
Eq. (\ref{integralcurr}) can be specialized to the axis-symmetric
case by integration of Eq. (\ref{ez1}) over the azimuthal angle
$\phi$ which leads to

\begin{equation}
\hat{E}_z = - i q^2 \int_0^1 d\hat{r}' \left[ \hat{r}'
\sum_{n=-\infty}^{\infty} \hat{j}_1^{(n)} G^{(n)} e^{-i n \phi}
\right] ~,\label{ez2b}
\end{equation}
where
\begin{equation}
G^{(n)}(\hat{r},\hat{r}')=\left\{
\begin{array}{c}
I_n(q \hat{r}) K_n(q \hat{r}') ~~~~ \hat{r}<\hat{r}'\\
I_n(q \hat{r}') K_n(q \hat{r}) ~~~~ \hat{r}>\hat{r}'
\end{array}
\right.~, \label{gikb}
\end{equation}
$I_n$ being the modified Bessel functions of the first kind of
order $n$. The equation for the $n$-th azimuthal harmonic of the
field can be written from Eq. (\ref{ez2b}) as

\begin{equation}
\hat{E}_z^{(n)}= - i q^2 \int_0^1 d\hat{r}' \hat{r}'
\hat{j}_1^{(n)} G^{(n)}~. \label{ez2n}
\end{equation}
Therefore, under the assumption of a cold beam, Eq.
(\ref{integralcurr}) can be rewritten as

\begin{eqnarray}
\hat{j}_1^{(n)} = \hat{a}_{1d}^{(n)} + i \hat{z}
\hat{a}_{1e}^{(n)} && \cr + q^2 S_0 \int_0^{\hat{z}} d\hat{z}
\left[ (\hat{z}'-\hat{z}) \int_0^1 d\hat{r}' \hat{r}'
\hat{j}_1^{(n)} G^{(n)}\right]~, \label{integrodiff2bibis}
\end{eqnarray}
which can be easily transformed, by double differentiation with
respect to $\hat{z}$ into the integro-differential equation:

\begin{equation}
{d^2 \hat{j}_1^{(n)} \over{d \hat{z}^2}} = - q^2 S_0 \int_0^1 d
\hat{r}' \hat{r}' G^{(n)} \hat{j}_1^{(n)}~. \label{differj}
\end{equation}
Eq. (\ref{differj}) is of course to be considered together with
proper initial conditions for $\hat{j}_1$ and its z-derivative at
$z=0$. The interval $(0,1)$ can be then divided into an arbitrary
number of parts so that Eq. (\ref{differj}) is transformed in a
system of the same number of 2nd order coupled differential
equations. The possibility of transforming Eq. (\ref{differj}) in
a system of 2nd order coupled differential equations explains the
choice of starting, in this case, with Eq.(\ref{integralcurr})
instead of Eq. (\ref{integrodiff}): in this way our system can be
solved straightforwardly by means of numerical techniques. To do
so we used a $4th$-order Runge-Kutta integration method, which
gave us the solution of the evolution problem in terms of the beam
current. Then, using Eq. (\ref{ez2n}) we could get back
$\hat{E}_z$ and we compared obtained results with Eq.
(\ref{finalend}) for different choices of transverse profiles. The
real field $E_z$ should be recovered, for any particular
situation, passing to the dimensional quantity $\tilde{E}_z$ and,
then, remembering $E_z = \tilde{E}_z e^{i\psi} + \tilde{E}^*_z
e^{-i\psi}$: yet, all relevant information is included in
$Re(\hat{E}_z)$. To give first a general idea of the obtained
result we present, in Figs. \ref{Step3D}, \ref{Para3D} and
\ref{Gauss3D}, $Re(\hat{E}_z)$ as a function of $\hat{z}$ and
$\hat{r}$ in the case of stepped, gaussian and parabolic profile
respectively, with parameters choice specified in the figure
captions. In all cases the initial conditions are proportional to
the transverse distribution functions (stepped, gaussian,
parabolic), $\hat{F}(\hat{P})=\delta(\hat{P})$ and $n=0$. We
consider only initial density modulation (i.e. $\hat{a}_{1e}=0$).
Note that, in order to be consistent with the perturbation theory
approach we should really choose $\hat{a}_{1d} \ll 1$, since it is
normalized to the bunch current density. However using, for
example, $\hat{a}_{1d} = \rho$ with $\rho \ll 1$ will simply
multiply our results by an inessential factor $\rho$ so, for
simplicity, we chose $\rho = 1$. Note the oscillatory behavior in
the $\hat{z}$ direction. Comparison with the Runge-Kutta
integration program are shown in Figs. \ref{cmpS}, \ref{cmpP} and
\ref{cmpG}. In the parabolic case, both pure analytical solution
and solution with multilayer approximation method are present,
while in the gaussian case only a solution with the multilayer
method is possible, to be compared with the numerical Runge-Kutta
result. This comparison shows that the assumption of the validity
of Eq. (\ref{expG1}) is correct in the parabolic case, and
validates it once more for the stepped profile situation.

\subsection{\label{sub:plots} The role of the initial condition}

Here we present some further exemplification of the obtained
results.

\begin{figure}
\begin{center}
\includegraphics[width=90mm]{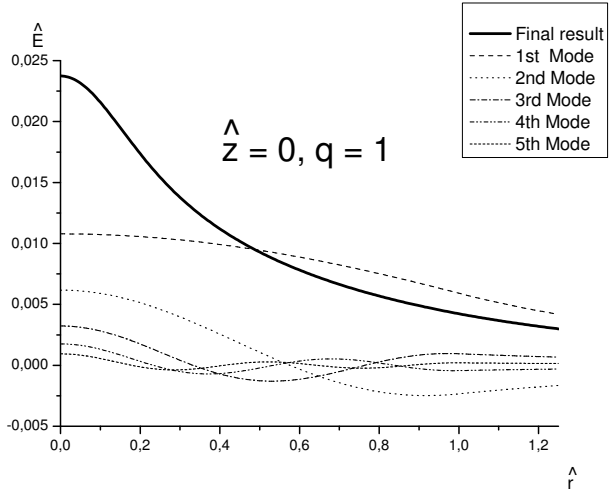}
\caption{\label{Q1pz0} The final result and the first five modes
$\hat{a}_{1d} = e^{-\hat{r}^2/(2 \sigma)}$ with $\sigma = 0.1$ at
$\hat{z}=0$ and for $q=1$. The final result is the sum of the
first fifty modes, here.}
\end{center}
\end{figure}
\begin{figure}
\begin{center}
\includegraphics[width=90mm]{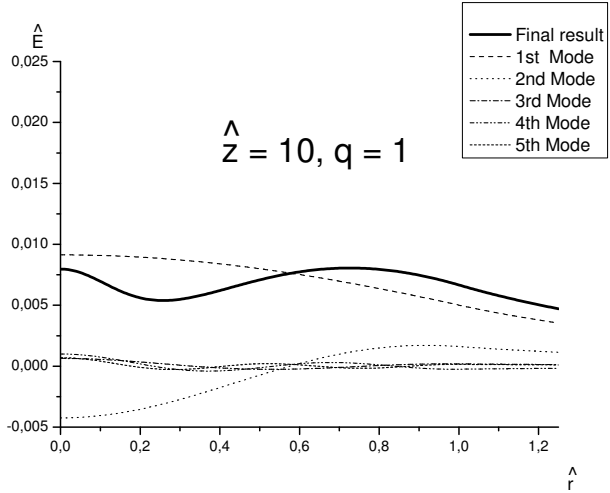}
\caption{\label{Q1pz10} The final result and the first five modes
$\hat{a}_{1d} = e^{-\hat{r}^2/(2 \sigma)}$ with $\sigma = 0.1$ at
$\hat{z}=10$ and for $q=1$. The final result is the sum of the
first fifty modes, here.}
\end{center}
\end{figure}

In particular we are interested in investigating, in a few cases,
what is the role of the initial condition in the final results, in
order to develop some common sense regarding our analytical
formulas.

The main parameter in our system is the transverse beam extent
$q$. However, intuitively, one can set two limiting initial
conditions: one in which only a small part of the transverse
section of the beam is modulated and the other in which all of it
is modulated. Depending on the profile of the initial modulation,
one can have excitation of many modes or only a few, while the
conditions for excitation of a single mode have been discussed in
Section \ref{sub:axis}. The transverse parameter $q$ will fix the
eigenvalue problem and, therefore, the oscillation wavelength (in
the $z$ direction) of the various modes. If $q$ is smaller than or
comparable to unity we expect to have appreciable differences in
the eigenvalues, which lead to a quick (in $\hat{z}$) change of
the relative phases between different modes. As a result the
initial shape of the fields will change pretty soon. On the other
hand, when $q$ is larger than unity, we will have all the
eigenvalues converging to unity as in the one-dimensional case,
which means that the relative phases between different modes will
stay fixed for a much longer interval in $ \hat{z}$ and the
initial shape of the fields will not change during the evolution.

\begin{figure}
\begin{center}
\includegraphics[width=90mm]{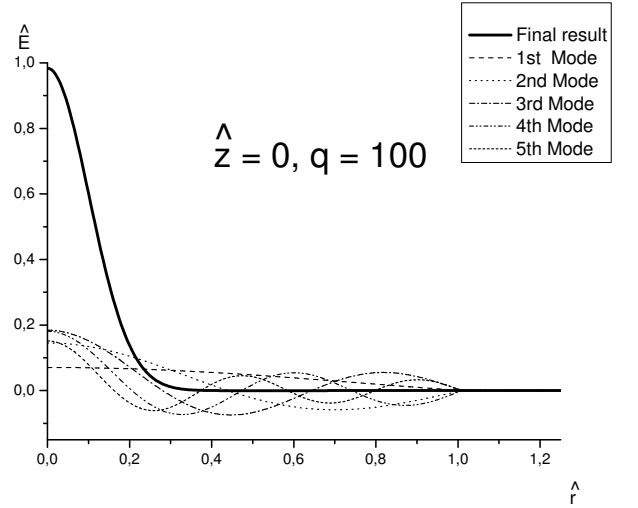}
\caption{\label{Q100pz0} The final result and the first five modes
$\hat{a}_{1d} = e^{-\hat{r}^2/(2 \sigma)}$ with $\sigma = 0.1$ at
$\hat{z}=0$ and for $q=100$. The final result is the sum of the
first fifty modes. Here $n=0$. }
\end{center}
\end{figure}
\begin{figure}
\begin{center}
\includegraphics[width=90mm]{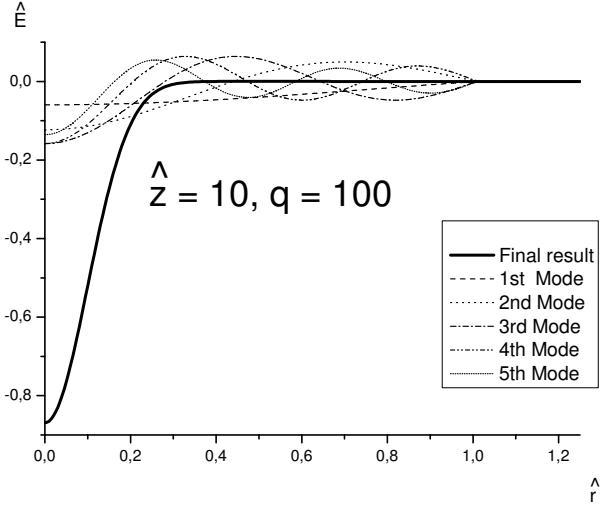}
\caption{\label{Q100pz10} The final result and the first five
modes $\hat{a}_{1d} = e^{-\hat{r}^2/(2 \sigma)}$ with $\sigma =
0.1$ at $\hat{z}=10$ and for $q=100$. The final result is the sum
of the first fifty modes. Here $n=0$.}
\end{center}
\end{figure}
To exemplify this situation we set up several calculations using
our analytical solutions to the initial value problem. In
particular we considered two cases $q=1$ and $q=100$, and a radial
stepped profile. Again, we consider only initial density
modulation (i.e. $\hat{a}_{1e}=0$) and we study two subcases: in
the first we set $\hat{a}_{1d} = e^{-\hat{r}^2/(2 \sigma)}$ with
$\sigma = 0.1$ and in the second we put $\hat{a}_{1d}=1$. As said
before, in order to be consistent with the perturbation theory
approach we should really choose $\hat{a}_{1d} \ll 1$, since it is
normalized to the bunch current density but using, for example,
$\hat{a}_{1d} = \rho$ with $\rho \ll 1$ will simply multiply our
results by an inessential factor $\rho$ so, for simplicity, we
chose, again, $\rho = 1$. Moreover we considered the azimuthal
harmonic $n=0$.

Figures \ref{Q1pz0} and \ref{Q1pz10} present the first five
eigenfunctions (with relative weights and phases) and the sum of
the first fifty (i.e. the final result) for $\hat{a}_{1d} =
e^{-\hat{r}^2/(2 \sigma)}$ with $\sigma = 0.1$ at $\hat{z}=0$ and
at $\hat{z}=10$ and for $q=1$. As one can see, the relative phases
have changed and the shape of the total field, our final result,
has also changed with $\hat{z}$.

\begin{figure}
\begin{center}
\includegraphics[width=90mm]{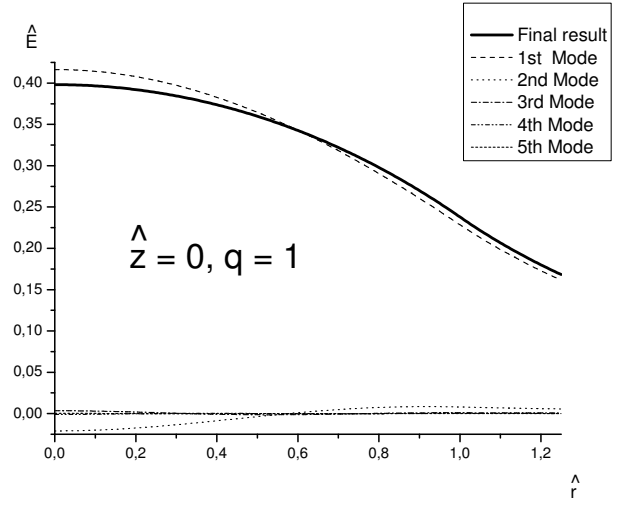}
\caption{\label{Q1fz0} The final result and the first five modes
$\hat{a}_{1d} = 1$ at $\hat{z}=0$ and for $q=1$. The final result
is the sum of the first fifty modes. Here $n=0$. }
\end{center}
\end{figure}
\begin{figure}
\begin{center}
\includegraphics[width=90mm]{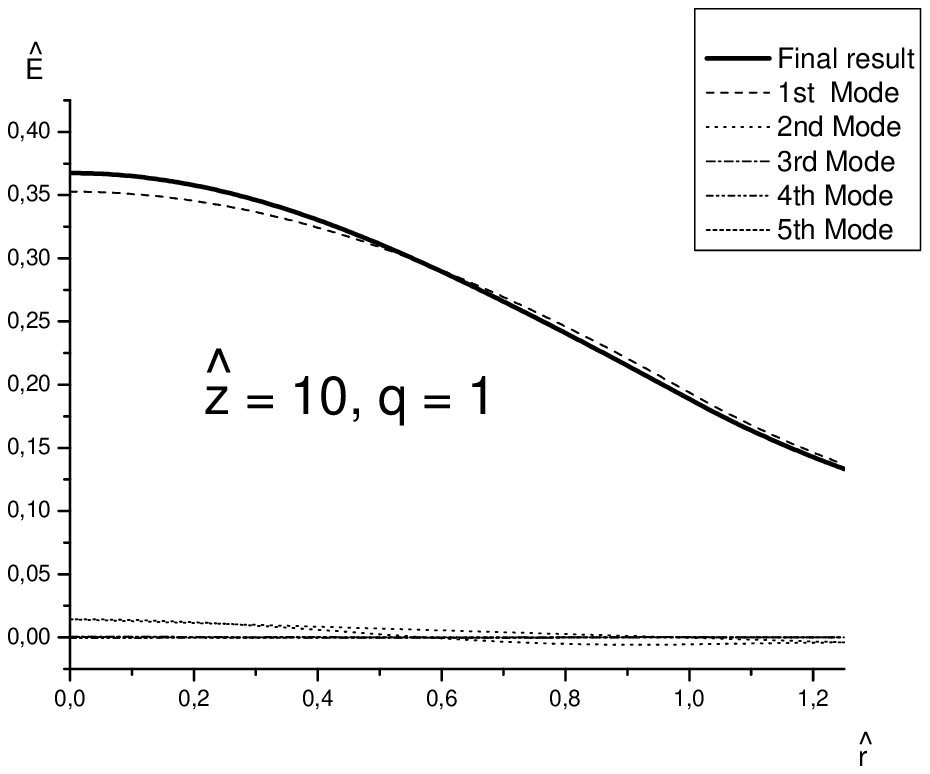}
\caption{\label{Q1fz10} The final result and the first five modes
$\hat{a}_{1d} = 1$ at $\hat{z}=10$ and for $q=100$. The final
result is the sum of the first fifty modes. Here $n=0$. }
\end{center}
\end{figure}
\begin{figure}
\begin{center}
\includegraphics[width=90mm]{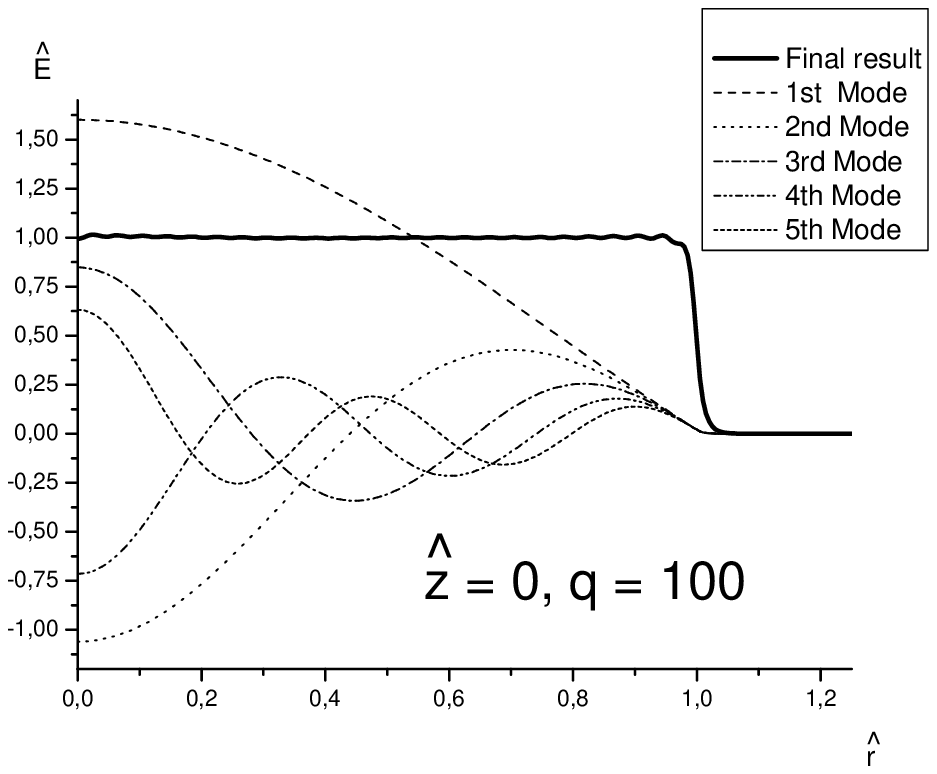}
\caption{\label{Q100fz0} The final result and the first five modes
$\hat{a}_{1d} = 1$ at $\hat{z}=0$ and for $q=100$. The final
result is the sum of the first fifty modes. Here $n=0$. }
\end{center}
\end{figure}
\begin{figure}
\begin{center}
\includegraphics[width=90mm]{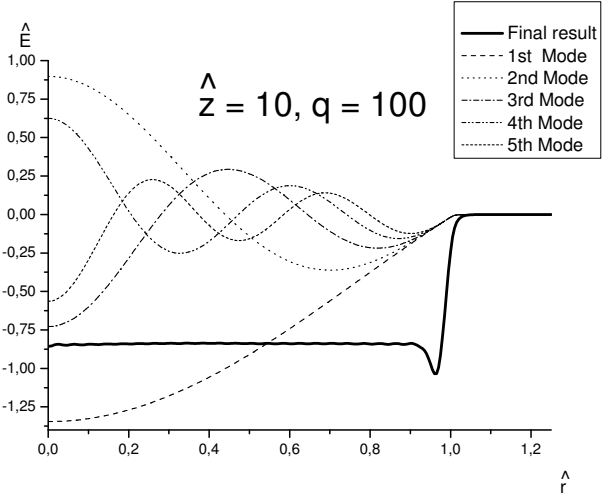}
\caption{\label{Q100fz10} The final result and the first five
modes $\hat{a}_{1d} = 1$ at $\hat{z}=10$ and for $q=100$. The
final result is the sum of the first fifty modes. Here $n=0$. }
\end{center}
\end{figure}
On the contrary, Figs. \ref{Q100pz0} and \ref{Q100pz10} present
the first five eigenfunctions (with relative weights and phases)
and the sum of the first fifty (i.e. the final result) for
$\hat{a}_{1d} = e^{-\hat{r}^2/(2 \sigma)}$ with $\sigma = 0.1$ at
$\hat{z}=0$ and at $\hat{z}=10$, and for $q=100$. Here the
relative phases have almost not changed and the shape of the total
field, our final result, is also remained unvaried (of course one
must account for the fact that the system is undergoing plasma
oscillation, so the shape, and not the field magnitude, is what is
important here).

\begin{figure}
\begin{center}
\includegraphics[width=90mm]{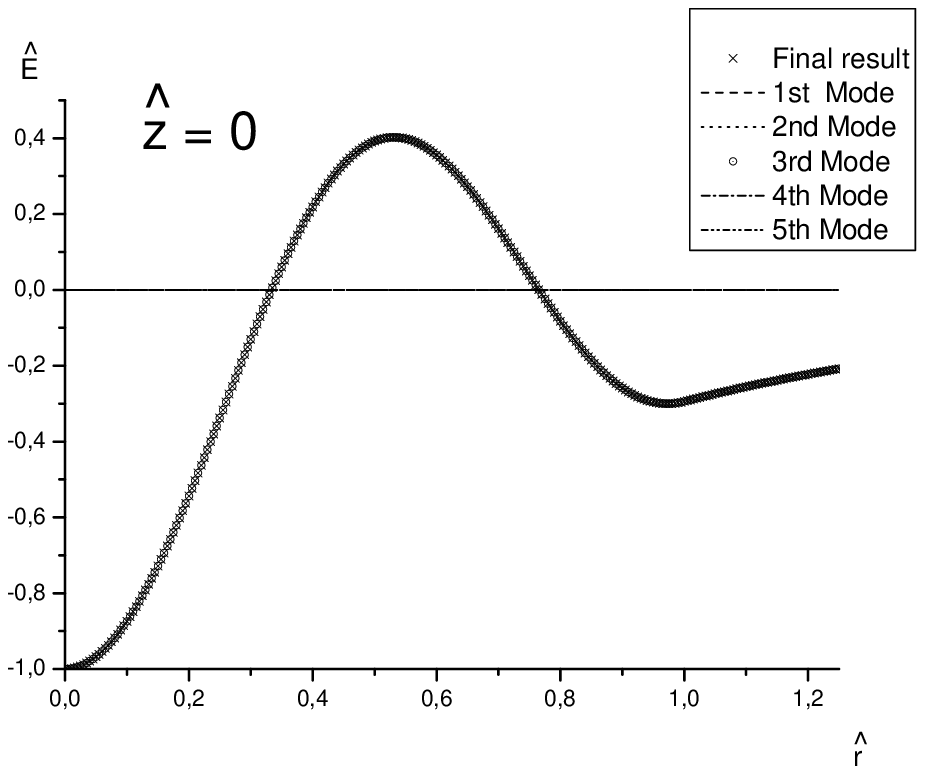}
\caption{\label{singlez0} The final result and the first five
modes when the initial condition is set in order to excite the
third mode alone. The final result is the sum of the first fifty
modes. Here $n=0$. }
\end{center}
\end{figure}
\begin{figure}
\begin{center}
\includegraphics[width=90mm]{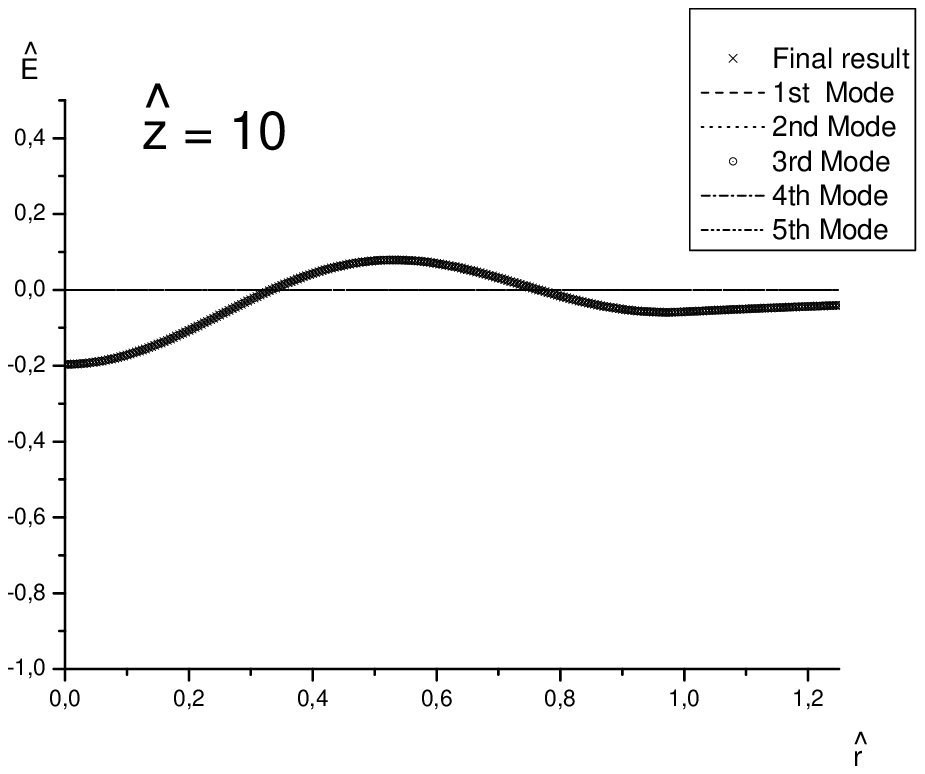}
\caption{\label{singlez10} The final result and the first five
modes when the initial condition is set in order to excite the
third mode alone. The final result is the sum of the first fifty
modes. Here $n=0$. }
\end{center}
\end{figure}
For comparison, it is interesting to plot analogous figures for
the second situation, that is $\hat{a}_{1d}=1$. Figs. \ref{Q1fz0}
and \ref{Q1fz10} depict the situation for $q=1$ at $\hat{z}=0$ and
$\hat{z}=10$ respectively. The way the phases behave is similar to
what has been seen before, i.e. there is a rapid change in the
relative phases between the modes, but now it is more difficult to
see from the plots because, in contrast with the case of
$\hat{a}_{1d} = e^{-\hat{r}^2/(2 \sigma)}$ already the first mode
is sufficient to fit the initial conditions relatively well so
that the field shape is almost unchanged. In Figs. \ref{Q100fz0}
and \ref{Q100fz10} we plot, instead, the case $q=100$ always as
$\hat{z}=0$ and $\hat{z}=10$ respectively, and with
$\hat{a}_{1d}=1$. Here it is easy to see, once more, that the
relative phases between different modes are almost unchanged. What
is of interest in this latter set of four pictures is the way
different modes are working together to satisfy initial
conditions, in comparison with the way they mix in the case for
$\hat{a}_{1d} = e^{-\hat{r}^2/(2 \sigma)}$: note, in particular,
how the first mode is almost enough to satisfy the initial
conditions (Figs. \ref{Q1fz0} and \ref{Q1fz10}), while in Figs.
\ref{Q1pz0} and \ref{Q1pz10} it is almost completely suppressed.
This should not be a surprise, considering that we may actually
select a single mode by fixing appropriate initial conditions as
described in Eq. (\ref{constr}) and Eq. (\ref{condi}). For
instance, if we fix $\hat{a}_{1e}=0$ and we want to excite only
the $j$th mode for a given value of the azimuthal harmonic $n$ in
the case $S_0 = 1$, then, according to Eq. (\ref{condi}) and Eq.
(\ref{stepfinal}) we must set (modulus a constant factor):

\begin{equation}
\hat{a}_{1d}= {\partial^2 J_n(\alpha_j \hat{r})\over{\partial
\hat{r}^2}} + {1\over{\hat{r}}}{\partial J_n(\alpha_j \hat{r}
)\over{\partial \hat{r}}} - \left(
q^2+{n^2\over{\hat{r}^2}}\right)J_n(\alpha_j\hat{r})
\label{inicondfirst}
\end{equation}
where $\alpha_j$ is defined in Eq. (\ref{conpsi2bis}). For example
if we choose $n=0$ and $j=2$ (third mode) for $q=1$ we obtain the
results presented in Fig. \ref{singlez0} at $\hat{z}=0$ and in
Fig. \ref{singlez10} at $\hat{z}=10$. As it can be seen by
inspection only the third mode is excited and evolves. Our
condition Eq. (\ref{inicondfirst}) set strictly to zero the
contributions of all the modes with $j\neq 2$. Of course, in
practice, actual data plotted in the figures show finite
contributions of the other modes ascribed to the finite accuracy
of our computations: to be precise, the difference between the
final result (sum of the first fifty modes) and the third mode
alone was found to be on the fourth significative digit.

\section{\label{sec:concl} Conclusions}

In this paper paper we presented one of the few self-consistent
analytical solutions for a system of charged particles under the
action of their own electromagnetic fields. Namely, we considered
a relativistic electron beam under the action of space-charge at
given initial conditions for energy and density modulation and we
developed a fully analytical, three-dimensional theory of plasma
oscillations in the direction of the beam motion.

We used the assumption of a small modulation so that we could
investigate the system behavior in terms of a linearized Vlasov
equation coupled with Maxwell equation, under the assumption that
field retardation effects can be neglected. Then we introduced
normalized quantities according to similarity techniques and we
provided two equivalent presentations for the evolution problem in
terms of a integrodifferential equation for the electric field and
of a integral equation for the beam current.

The integrodifferential equation for the fields was particularly
suited to be solved with the help of Laplace transform techniques:
we did so in all generality and we discussed the mathematical
difficulties involved in the general treatment, namely the
assumption of a well-behaved differential operator allowing
eigenmodes expansion of the Green function and the problem of the
analytic continuation of the Laplace transform of the field to all
the complex plane (isolated singularities excluded), in relation
with the application of the Fourier-Mellin integral to
antitransform $\bar{E}$. Our considerations led us to restrict our
attention to the cold beam case. We specialized the general method
to the important cases of stepped and parabolic transverse
profiles, which are among the few analytically solvable
situations. In particular, the stepped profile case could be used
to develop a semi-analytical technique to solve the evolution
problem for the field using an arbitrary transverse shape.

We tested our results by discussing the limit for the 1-D theory
($q \rightarrow \infty$). We also developed an algorithm able to
solve the evolution problem in terms of the beam currents. The
integral equation for the currents could be easily approximated to
a system of second order ordinary differential equations which
could be solved by means of numerical Runge-Kutta integration
method. Once the solution for the current was known we recovered
the electric field evolution by integration of the current with a
suitable Green function. Numerical and analytical or
semi-analytical solutions for the fields were then compared and
gave a perfect agreement. In this way we could state that the
assumption of the correctness of the eigenmodes expansion for the
Green function has been proved, for some particular profiles, by
means of numerical crosschecks (in the stepped profile case, by
means of alternative analytical techniques too).

Finally we exemplified the role of the initial condition, which we
have seen to control the way one ore more modes interact together
to give the final result. In particular we have shown how to build
up initial conditions in such a way that a single mode is excited
and propagates through. We checked our prescription by setting up
particular initial conditions and looking at the propagation of
various eigenmodes.

In conclusion we proposed, checked and analyzed, both from
physical and mathematical viewpoint, a theory of space-charge
waves on gradient-profile relativistic electron beams. This work
is of fundamental importance, since it is one of the few known
analytical solution to evolution problems for systems of particles
and fields. In particular, today, it is of great relevance in the
physics of FEL and high-brightness linear particle accelerators.

\section{\label{sec:ackno} Acknowledgements}

We thank Reinhard Brinkmann (DESY), Martin Dohlus (DESY), Michele
Correggi (SISSA), Klaus Floettman (DESY) and Helmut Mais (DESY)
for useful discussions. We thank Jochan Schneider (DESY) and
Marnix van der Wiel (TUE) for their interest in this work.

\end{document}